\documentclass[fleqn,usenatbib]{mnras}

\usepackage{newtxtext,newtxmath}
\usepackage[T1]{fontenc}
\usepackage{graphicx}
\usepackage{amsmath}

\title[Simulations of the dissociative cluster A2034]{Simulations of the merging galaxy cluster Abell 2034: what determines the level of separation between gas and dark matter}

\author[M. T. Moura et al.]{%
Micheli T. Moura$^{1}$\thanks{E-mail: micheli\_trindade.moura@hotmail.com},
Rubens E. G. Machado$^{1}$ and Rog\'erio Monteiro-Oliveira$^{2,3}$ \\
$^{1}$Departamento Acad\^emico de F\'isica, Universidade Tecnol\'ogica Federal do Paran\'a, Rua Sete de Setembro 3165,80230-901 Curitiba, Brazil\\
$^{2}$Universidade de S\~ao Paulo, Inst. de Astronomia, Geof\'isica e Ci\^encias Atmosf\'ericas, Depto. de Astronomia, R. do Mat\~ao 1226, 05508-090 S\~ao Paulo, Brazil\\
$^{3}$Universidade Estadual de Santa Cruz, Laborat\'orio de Astrof\'isica Te\'orica e Observacional - 45650-000, Ilh\'eus-BA, Brazil\\
}

\date{Accepted 2020 October 27. Received 2020 October 22; in original form 2020 September 10}

\pubyear{2020}

\begin{document}

\label{firstpage}
\pagerange{\pageref{firstpage}--\pageref{lastpage}}
\maketitle

\begin{abstract}
Cluster mergers are an important laboratory for studying the behaviour of dark matter (DM) and intracluster gas. There are dissociative collisions that can separate the intracluster gas from the DM. Abell 2034 presents clear dissociative features observed by X-rays and gravitational lensing. The cluster, at $z$ = 0.114, consists of two substructures with mass ratio of about 1:2.2, separated by $\sim$720 kpc. The X-ray emission peak is offcentred from the south DM peak by $\sim$350 kpc.
Using $N$-body hydrodynamical simulations, we aim to reconstruct the dynamic history of the collision, reproducing the observed features, and also to explore
the conditions that led to the dissociation. Our best model assuming that the collision is close to the plane of the sky, with a small impact parameter, observed
0.26 Gyr after central passage, reproduces the observed features of this cluster, such as the offset between X-ray and DM peaks, X-ray morphology and temperatures.
We explored several variations using different gas and DM concentrations for each cluster. The level of dissociation was quantified by the distances between X-ray and DM peaks, and also by the gas retention in the cluster cores. We found that the ratio of central gas densities is more important than the ratio of central DM densities in determining the level of dissociation.

\end{abstract}


\section{Introduction}

Clusters mergers are an important laboratory for the study of structures in large scales, where they are an expected outcome of the hierarchical model of formation.
Interactions between individual clusters allow a more in-depth study of the effects generated on galaxies, intracluster medium (ICM) and dark matter (DM) halo.

Mergers are a source of extremely energetic cosmic rays \citep[e.g.][]{2010Vanweeren} and generate relevant disturbances in gas morphology \citep[e.g.][]{2007Markevitch}, whose details can be obtained by X-ray observations. 
Analyses of intracluster gas combined with the mass maps from gravitational lensing method and others can provide information about the major components in galaxy clusters.

Depending on the collision parameters, the peak of X-ray emission may not coincide with the total mass peaks of the system. This is a feature of a dissociative collision. During the merger, dark matter interacts mainly via the gravitational force, while the intracluster gas is the component that undergoes pressure effects and changes its morphology considerably. Different collision parameters such as mass ratio, impact parameter, infall velocity and others, cause different effects on the configuration of a dissociative merger.


There has been an increasing number of observed dissociative systems since the discovery of the Bullet Cluster \citep{CloweBradac2006}, providing a range of features from particular dissociative systems. Some examples include: Abell 520 \citep{Mahdavi2007}, MACS J0025.4-1222 \citep{Brada2008}, Abell 2163 \citep{2011okabe2163}, Abell 2744 \citep{Merten2011}, Abell 1758N \citep{Ragozzine2012}, DLSCL J0916.2+2951 \citep{Dawson2012}, ZwCl 1234.0+02916 \citep{2013ADahle}, ACT-CL J0102-4915 \citep{Jee2014}, SL2S J08544-0121 \citep{2014GastaldelloSL2S}, Abell 4067 \citep{2015Chon4067}, CIZA J2242.8+5301-I \citep{2015JeeLinguica}, RX J0603.3+4214 \citep{2016ApJeetoothblush}, J1149.5+2223 \citep{2016AGovovichelgordo2}, ZwCl 0008.8+5215 \citep{2017ApJGolovichelmagro}, Abell 2256 \citep{2020Breuer2256}, Abell 2399 \citep{2399ana}.

Due to the high density of dark matter in galaxy clusters, dissociative collisions work as astrophysical particle colliders, providing an excellent laboratory to investigate possible interactions of dark matter beyond the standard model \citep{Bauer2015}. Theoretical models of dark matter self-interaction predict small, but detectable offsets between visible and dark cluster components just after the pericentric passage. These detachments are sometimes referred to as `cluster bulleticity' and consists of an observational signature that can be translated into the dark matter self-interaction cross-section \citep[$\sigma/m$;][]{Massey2011,Harvey2015,Drlica-Wagner2019}.

Numerical simulations have proven to be an important tool to study these extreme events. Cluster mergers can be studied by theoretical explorations of the parameter space \cite[e.g.][]{ZuHone2011,2017_Kim}. In this context, \cite{2006PooleI} investigated the effects of mergers in relaxed clusters, as well as the merger effects on X-rays \citep{2007_poole2}. Moreover, there are simulations that aim to reproduce specific properties of observed collision, such as sloshing spiral \cite[e.g.][]{2015_2056Machado, 2020_Lia1644}. Specifically about dissociative mergers, some numerical simulations have been performed aiming to reproduce such properties. 
Some examples of dedicated simulations include: the Bullet Cluster 1E0657-56, \citep[e.g.][]{2014LageeFarrar, 2008MastropietroeBurket,Springel2007}, the `El Gordo' cluster \citep[e.g.][]{2015AZhangetal,2015MonlareBroad,2014Donnert}, the `Sausage' cluster \citep[e.g.][]{2017Donnert,2017MolnareBroad}, A1758N \citep{Machado2015b, Monteiro2017a} and ZwCl008.8+52 \citep{Molnar2018}.

Abell 2034 (A2034) is located at $z$ = 0.114 and is composed of two substructures: A2034N and A2034S. The two brightest cluster galaxies (BGC) are separated by $\sim$5 arcmin. The X-ray morphology presents a unimodal distribution, offcentred $\sim$91 arcsec from the south BCG \citep{MonteiroOliveira2018}. Evidence of a recent merger between the substructures has been presented by \cite{2001_kempner2001A2034,Kempner2003,Owers2014,MonteiroOliveira2018,Golovich2019}.
A2034 has been observed by ROSAT \citep{David1999}, ASCA \citep{White2000}, XMM-\textit{Newton} \citep{Okabe2008} and \textit{Chandra} \citep{Kempner2003,Owers2014}. The X-ray features were explored by \cite{Owers2014}, indicating an edge caused by a temperature discontinuity near to the north BCG. Additionally, the cluster presents complex diffuse radio emission, explored in detail by \cite{2016Shimwell}. Mass reconstructions were performed by \cite{Geller2013,Owers2014,2015Ledelliou} and more recently by \cite{MonteiroOliveira2018} using the weak gravitational lensing method. The merger scenario suggests small impact parameter in a collision that occurred close to the plane of the sky, where the gas from the northern cluster was lost during the interaction with the southern substructure \citep{Owers2014,MonteiroOliveira2018}.

In this work we aim to reproduce the observed features of A2034 and investigate the factors that determine this dissociative case. More specifically, we perform hydrodynamical $N$-body simulations suitable to reproduce the dynamical history of A2034. We also carry out a theoretical exploration of the effect that different gas and dark matter concentrations cause in the dissociation of the system.

In section \ref{sec:simsetup} we will present the simulation setup. The results are presented in section \ref{results}, where we obtain a model that best reproduces A2034, compare it to observations, and also explore general concentrations to the theoretical discussion about the different merger scenarios for this system. In section \ref{sec:conclusions} and \ref{sec:summary} we discuss and summarise the results. In this paper we assume a standard $\rm \Lambda CDM$ cosmology with $\Omega_{M} = 0.3$, $\Omega_{\Lambda} = 0.7$ and $ H_{0} = 70$ km $\rm s^{-1}$ Mpc$^{-1}$. Given the A2034 redshift ($z$ = 0.114), 1 arcsec = 2.06 kpc for the adopted cosmology.

\section{Simulation setup}
\label{sec:simsetup}

\begin{table}
	\centering
	\caption{Exploration of collision parameters space. The columns indicates the parameter that is varied, keeping the others fixed in the best combination. The bold values indicates the parameters of the best simulated model.}
	\label{tab:tab_param}
	\begin{tabular}{ccc}
		\hline
		$b$ & $v_0$ & $i$\\
		(kpc) & (km/s) & ($^\circ$)\\
		\hline
		\textbf{0} & \textbf{2000} & \textbf{0}\\
		150 & 2000 & 0\\
		250 & 2000 & 0\\[0.5em]
		0 & 1500 & 0\\
		\textbf{0} & \textbf{2000} & \textbf{0}\\
		0 & 2500 & 0\\[0.5em]
		\textbf{0} & \textbf{2000} & \textbf{0}\\
		0 & 2000 & 13\\
		0 & 2000 & 27\\
		0 & 2000 & 41\\
		\hline
	\end{tabular}
\end{table}

\begin{table}
    \begin{center}
    \centering
	\caption{Exploration of the gas and DM scale lengths for the different models. The columns indicate the model label and the values of each scale length respectively. The model `0' indicates the default scale length model for the north and south subclusters.}
	\label{tab:tab_scale}
	\begin{tabular}{cccccc}
	    \hline
		 Model & $a_{\rm g N}$  & $a_{\rm g S}$  & $a_{\rm h N}$ & $a_{\rm h S}$ & Comment\\
		 (label) & (kpc) & (kpc) & (kpc) & (kpc) & \\
		 \hline
		 1 & 200 & 400 & 300 & 400 & More concentrated N gas\\
		 0 & 300 & 400 & 300 & 400 & Default \\
		 2 & 400 & 400 & 300 & 400 & Less concentrated N gas \\[0.5em]
		 3 & 300 & 300 & 300 & 400 & More concentrated S gas\\
		 0 & 300 & 400 & 300 & 400 & Default \\
		 4 & 300 & 500 & 300 & 400 & Less concentrated S gas \\[0.5em]
		 5 & 300 & 400 & 200 & 400 & More concentrated N halo \\
		 0 & 300 & 400 & 300 & 400 & Default \\
		 6 & 300 & 400 & 400 & 400 & Less concentrated N halo \\[0.5em]
		 7 & 300 & 400 & 300 & 300 & More concentrated S halo \\
		 0 & 300 & 400 & 300 & 400 & Default \\
		 8 & 300 & 400 & 300 & 500 & Less concentrated S halo \\
		\hline
	\end{tabular}
	\end{center}
\end{table}

We aim to investigate the collision scenario of Abell 2034 (A2034N and A2034S) using $N$-body hydrodynamic simulations. To simulate the dissociative collision, we set up two spherical galaxy clusters, both comprised of dark matter halo and adiabatic gas. Galaxies and star formation are not considered. We develop idealised initial conditions suitable for the systematic study of the local dynamics of binary cluster mergers, disregarding the cosmological expansion. We employ the \textsc{Gadget-2} code \citep{Springel2005}, which uses smoothed particle hydrodynamic (SPH), with softening length of $\epsilon$ = 5 kpc to run the simulations. In this context, dark matter particles interact only gravitationally (i.e. it is assumed $\sigma/m=0$). To analyse the simulation output, we use tools from the yt-project \citep{2011Turk}.

\begin{figure*}
	\includegraphics[width=\textwidth]{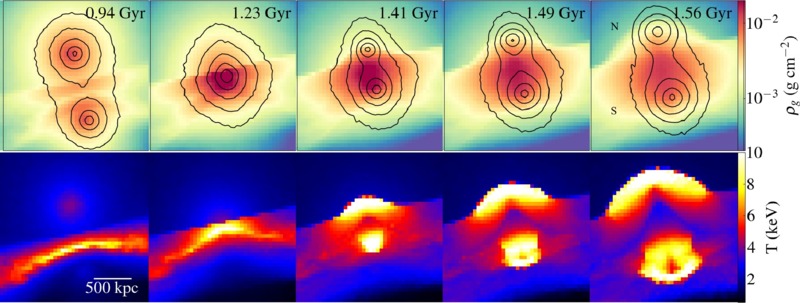}
    \caption{Time evolution of the gas density and emission-weighted temperature in the north and south substructures. The best epoch of the simulation occurs at $t = 1.49$ Gyr of simulation, which is 0.26 Gyr after the central passage. The contours lines represent the total projected mass. The `N' and `S' indicate the orientation of substructures.}
    \label{fig:evoltemp2.jpg}
\end{figure*}

To set up two spherical galaxy clusters with dark matter and gas particles, we assume a \cite{Hernquist1990} profile for dark matter halo density, which is similar to the  NFW profile \citep{1997NFW}:
\begin{equation}
       \rho_{\rm h}(r) = \frac {M_{\rm h}}{2\pi}~\frac{a_{\rm h}}{r(r+a_{\rm h})^{3}}, 
\end{equation} 
where $M_{\rm h}$ is the total dark matter mass, and $a_{\rm h}$ is the halo scale length. The Hernquist profile has finite mass and resembles the NFW profile in internal regions.

We assume a \cite{Dehnen1993} profile for gas density distribution:
\begin{equation}
       \rho_{\rm g}(r) = \frac {(3-\gamma) M_{\rm g}}{4\pi}~\frac{a_{\rm g}}{r^\gamma(r+a_{\rm g})^{4-\gamma}},
\end{equation}
where $M_{\rm g}$ is the total gas mass and $a_{\rm g}$ is the gas scale length. When $\gamma = 0$, the profile becomes similar to the $\beta$-model \citep{fusco}.
The gas temperature profile is obtained from the assumption of hydrostatic equilibrium.
The velocities of the gas particles are thus zero by construction, while the velocities of the dark matter particles are obtained by sampling the distribution function itself, via Eddington's formula, assuming only spherical symmetry and isotropy. The realization of the initial conditions is made according to the procedures described in \cite{Machado2013}.
Our initial conditions are composed of $10^6$ gas particles and $10^6$ dark matter particles for each cluster. This implies mass resolutions of $(2.4 - 5.0) \times 10^{7}\,{\rm M}_{\odot}$ for the gas particles, and $(1.3 - 2.6) \times 10^{8}\,{\rm M}_{\odot}$ for the dark matter particles. The simulations are evolved over time for 2 Gyr in a non-cosmological setting.

From the observational data of A2034, we have a series of features to be reproduced numerically. Some of the features are the DM and gas offset, at the moment when the projected separation between the north and south subclusters is approximately 720 kpc. Simultaneously, the simulation is expected to produce a similar X-ray emission morphology and observed temperature.

Given these various constraints, we created initial conditions with different parameters, aiming to reproduce those properties. The clusters were created with masses similar to the virial masses obtained from gravitational weak lensing: $M^{\rm S}_{200}=2.35\times 10^{14}\,{\rm M}_{\odot}$ and $M^{\rm N}_{200} = 1.08\times 10^{14}\, {\rm M}_{\odot}$ \citep{MonteiroOliveira2018}. The initial conditions have a baryon fraction of 15 per cent.

From the mass and redshift, \cite{Duffy2008} provides a relationship of the expected concentration. The scale length of the NFW profile is related to the concentration and to the scale length of the Hernquist density profile \cite[e.g.][]{2005SpringeletalEQ3}:
\begin{equation}
       a_{\rm h} = r_{\rm s}\sqrt{2[{\rm ln}(1+c)-c/(1+c)]},
\end{equation} 
where the concentration $c$ is defined as $c = r_{\rm 200}/r_{\rm s}$, the $r_{\rm s}$ is the scale length of the NFW halo and $r_{\rm 200}$ is the virial radius. From the expected concentration ($c$ $\sim$ 4) obtained from the virial mass and radius, we adopt the `default' scale lengths: $a_{\rm g}$ =  $a_{\rm h}$ = 300 kpc for A2034N and  $a_{\rm g}$ =  $a_{\rm h}$ = 400 kpc for A2034S.

The scale lengths having been fixed, we explored the parameter space, limiting the search to plausible regimes, that is, regimes that satisfy the observational constraints. It is worth mentioning that the model named as `best model' refers to the model found via the exploration of the parameter space in a wide range for all the parameters explored. The impact parameter exploration was limited to only a few hundred kpc, because non-frontal encounters are likely to give rise to gas sloshing and pronounced asymmetries \cite[e.g.][]{2015_2056Machado, 2020_Lia1644}, which are not observed in this cluster. The ranges of explored velocities were motivated by the free fall velocity (of approximately 1000\,km\,s$^{-1}$) and also by the cosmologically expected velocity distribution \citep[e.g.][]{2006HayashiEwhite}. Through trial and error we refined the search in order to arrive at a suitable range for each varied parameter and thus find the best parameters that make up the best model.

Considering that exploration of the parameters space in simulations is considerably broad, Table~\ref{tab:tab_param} presents a small sample of the varied parameters, which will be covered in section \ref{sec:bestmodel}. We have performed a set of simulations with theoretical explorations of this dissociative case: Table~\ref{tab:tab_scale} presents the different setups for these investigations that will be discussed in section \ref{sec:dissociative}.

\section{Results}
\label{results}

The results are divided into three subsections. Sections \ref{sec:bestmodel} and \ref{sec:comparisonobs} will be dedicated to presenting and discussing the results for the dynamics of the system that best corresponds to the A2034, and comparing it to observations. Section \ref{sec:dissociative} is dedicated to discussing the dissociative features of the models explored and their correlations.

\subsection{Best Model for A2034}
\label{sec:bestmodel}
Reproducing observed properties using simulations requires a series of constraints that must be satisfied simultaneously.
For the A2034 case, the proper gas morphology must be obtained at the moment when the separation between the mass peaks is $\sim$720 kpc. At that moment, the distance between the X-ray emission peak and the south dark matter peak should be around $348 ^{\rm +98}_{-86}$ kpc \citep{MonteiroOliveira2018}. At the same time, the temperature range of the shock fronts are roughly 10--12 keV \citep{Owers2014}.

Fig.~\ref{fig:evoltemp2.jpg} presents the best model for the dynamics of this system, through five simulation times in different stages of interaction between both substructures. The snapshots were rotated to match the position angle of the observed cluster. We can notice the evolution of the collision and the effects on the gas through the projected density and emission-weighted temperature maps. The simulation starts at $t = 0$\,Gyr with an initial separation of 3000 kpc along the $x$-axis, between the south and north substructure. With the evolution of time, the clusters get closer, the north cluster reaches the south. At the same time, shock fronts develop with the clusters approach. The interaction increases until the central passage at $t = 1.23$\,Gyr. Next, the north cluster moves away gradually, achieving the best simulation time at $t$ = 1.49 Gyr. In this moment -- which is 0.26\,Gyr after central passage --  the separation reaches $\sim$720 kpc and the shock front in the north approximately coincide with the north dark matter peak.

The preferred model was obtained via trial and error until the observational constraints were simultaneously met. In order to arrive at this model, different combinations of initial parameters were attempted. After the best model was chosen, we built a small set of simulations around this model. These comparisons are presented to justify our choice of the best model, and also to illustrate a representative sample of the parameter space around it. Here we present variations of i) impact parameter $b$, ii) initial velocity $v_0$ and iii) inclination $i$.

\begin{figure}
	\includegraphics[width=\columnwidth]{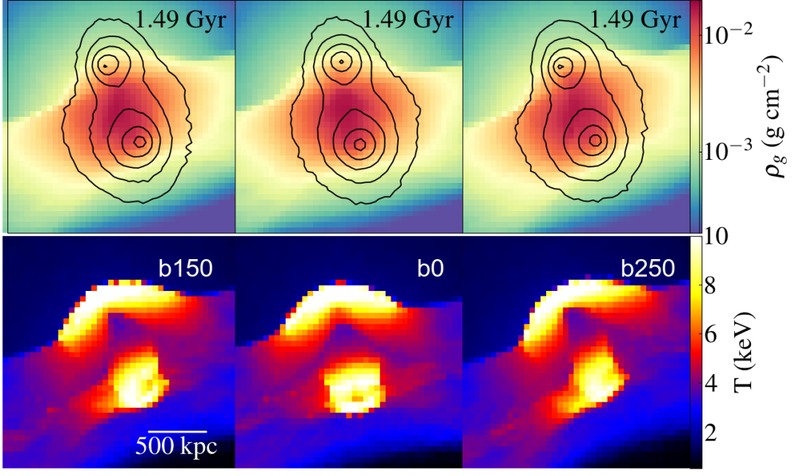}
    \caption{Impact parameters variation in the best epoch of simulation, from maps of projected density with total mass contours in first row and emission-weighted temperature in second row.}
    \label{fig:figB}
\end{figure}
\begin{figure}
	\includegraphics[width=\columnwidth]{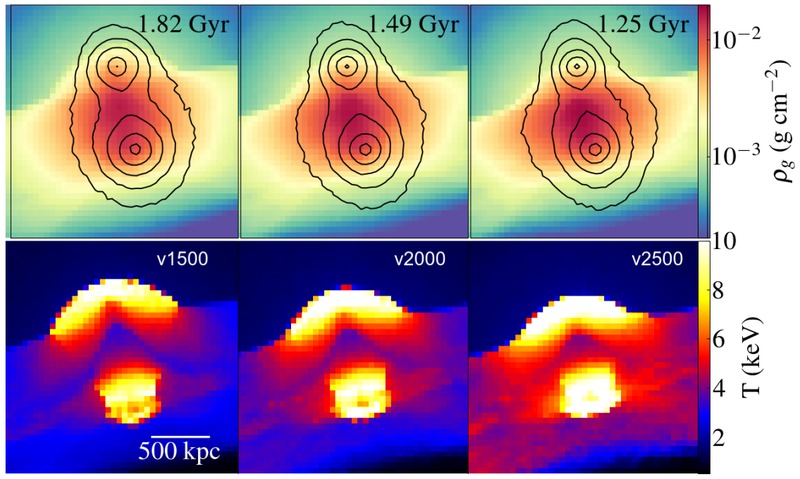}
    \caption{Projected density and emission-weighted temperature maps for three velocities at different times after central passage.}
    \label{fig:figV}
\end{figure}


i) Impact parameter $b$: In the initial conditions, we can employ different impact parameters that are displacements along the $y$-axis in the collision plane at $t = 0$  Gyr. Therefore, for nonzero impact parameters there is a minimum distance in the pericentric passage, and in turn, if the $b = 0$ kpc, the collision is frontal with a central passage. Different impact parameters generate different disturbances in the gas morphology. Fig.~\ref{fig:figB} shows different results for morphologies given some values of $b$, displayed by the density and temperature maps. We explored impact parameters from 0 to 500 kpc, in steps of 50 kpc (not shown). From this systematic exploration, we found that models with $b$>300 kpc can be ruled out because they are excessively asymmetric. Additionally, the temperatures of their shock fronts are too low. Given that even small displacements generate relevant asymmetries, impact parameters close to zero seem preferable to model the dynamics of this collision system, since that the observational configuration does not suggest major asymmetries. For these reasons, we adopt $b$ = 0 kpc for our default model and we propose 0--250 kpc as an acceptable range for this parameter.

ii) Initial velocity $v_0$: In our simulations, after creating the clusters with mass similar to the virial mass, with their respective gas and DM scale lengths, we determined the initial separation on the $x$ and $y$-axis, and an initial collision velocity.
Different initial velocities have different effects on gas morphology, density and temperature. Fig.~\ref{fig:figV} shows the initial velocity range explored and the effects on the gas density and temperature at the moment when separation between mass peaks is $\sim$720 kpc. We explored initial velocities from 500 to 3000 km/s in steps of 250 km/s (not shown). We found that in the range of 500 km/s to 1250 km/s the temperatures of the shock fronts barely reach 10 keV, whereas the observed temperature observed for the shock front is $\sim$12 keV \citep{Owers2014}. On the other hand, velocities above 2500 km/s start to exceed the expected temperature range. Considering these constraints, we adopted $v_0 = 2000$\,km/s as the default initial velocity for the evolution of the collision, and we offer the interval of 1500--2500 km/s as a plausible range for this parameter.

\begin{figure}
	\includegraphics[width=\columnwidth]{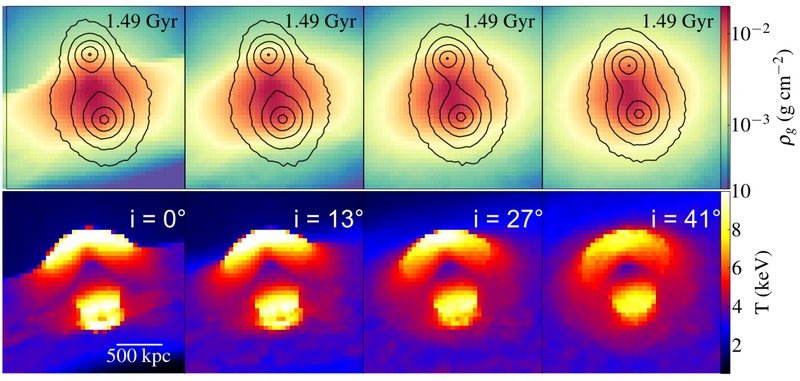}
	\caption{Range of inclinations in the plane of sky in the best epoch of simulation, from maps of density projected with total mass contours and emission-weighted temperature.}
	\label{fig:inclinacao.jpg}
\end{figure}

iii) Inclination $i$: The two-body dynamical model of \cite{MonteiroOliveira2018} suggested that the collision occurred not so far from the plane of sky, with an angle of $27^{\circ} \pm 14^{\circ}$. In order to explore the implications of different inclinations in the system, we present the effects of inclination in Fig.~\ref{fig:inclinacao.jpg}. 
We applied three non-zero inclinations within the range proposed by \cite{MonteiroOliveira2018}. Small inclinations are equally consistent to reproduce the configuration of the system at the moment of appropriate separation. Since there was no substantial improvement in the morphology by applying inclinations up to  $\sim27^{\circ}$, we adopt $i=0 ^{\circ}$ for simplicity.

It is important to emphasise that we are adopting a model that, within a reasonable approximation, reproduces several of the constraints of the system at a given time and separation. This idealised model captures the overall dynamics and the global morphological features. However, given the infinity of parameters, it may not be the only suitable model.

\subsection{Comparison to observations}
\label{sec:comparisonobs}

\begin{figure}
	\includegraphics[width=\columnwidth]{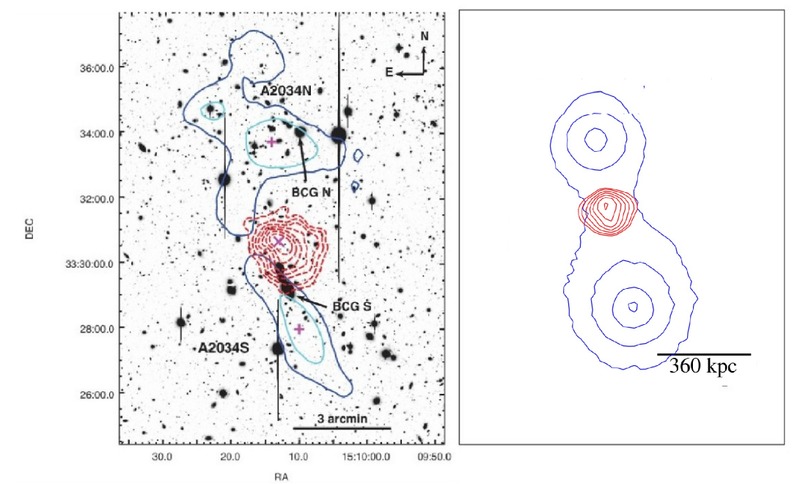}
    \caption{Left panel: The cyan and blue contour curves represent the observed position of the total mass peak with confidence levels of 1 and 2$\sigma$, determined by analysis of weak gravitational lensing (figure from \citealt{MonteiroOliveira2018}). The red contours represent the X-ray emission from \textit{Chandra}. Right panel: The blue contour curves represent the total mass and the red contour represent the gas density, based on the simulation of the best model.}
    \label{fig:rog.jpg}
\end{figure}

Having established the best model, we will detail the observational constraints that were reproduced by comparing it to the observational results. Fig.~\ref{fig:rog.jpg} (left) shows the observed configuration of A2034, where the blue curves represent the total mass peaks, obtained through the weak gravitational lensing method \citep{MonteiroOliveira2018}, and the red curves show the position of the X-ray emission peak, observed by \textit{Chandra}.

At the moment when the simulated clusters reach the desired distance, the position of the X-ray gas emission is shifted in relation to the total mass peaks, indicating the dissociation between these components. In the right panel of Fig.~\ref{fig:rog.jpg}, the simulation of the best model for A2034 is shown. The blue curves represent the total mass peaks and the red contour represents the simulated gas density. The orientation of the simulated subclusters is in accordance with the observed configuration. According to this simulated model, we observe the system at 0.26 Gyr after the central passage, at the moment when the distance between the total mass peaks coincides with what is observed. It is noticeable the distribution of gas between the simulated total mass peaks, where the separation between the X-ray centroid emission is displaced in relation to the south DM peak at 280 kpc, compatible with the observed interval.

\begin{figure}
    \centering
	\includegraphics[width=\columnwidth]{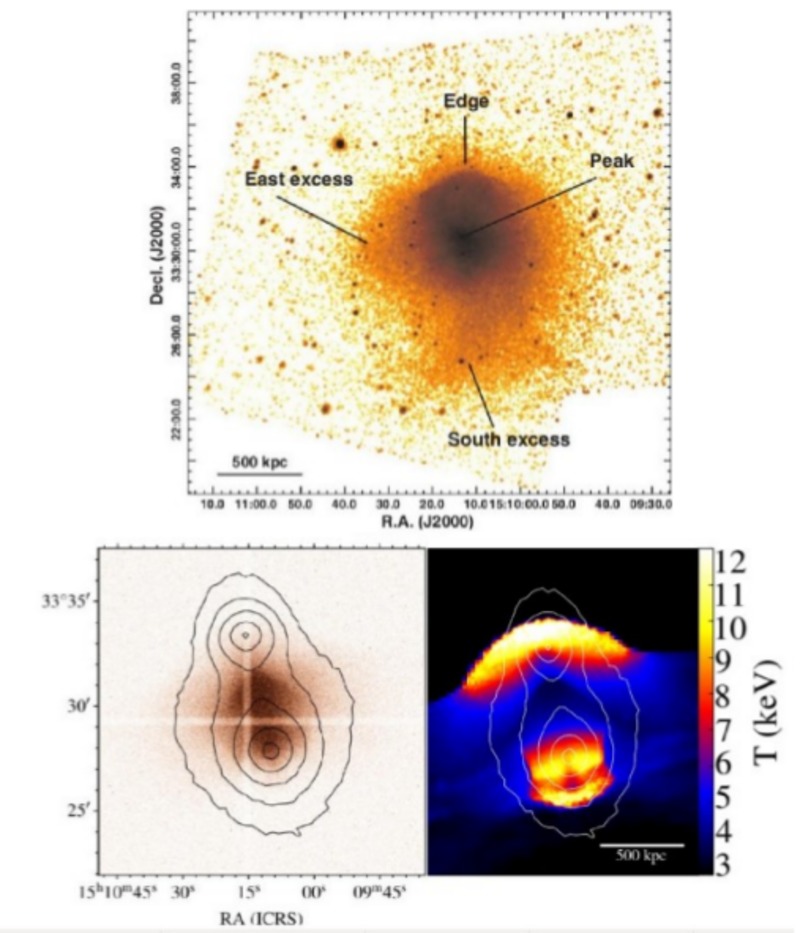}
	\caption{Observed (top) and simulated (bottom) gas morphology of the A2034. Top: Gas morphology through combined, exposure-corrected, and background-subtracted \textit{Chandra} image in the 0.5--7.0 keV energy band (figure from \citealt{Owers2014}). Bottom left: Mock X-ray image of the best simulated model. The contour curves represent total mass and the color represent counts/pixel. Bottom right: Emission-weighted temperature map. The contour curves represent total mass curves, the colors represent temperature.}
    \label{fig:owers.jpg}
\end{figure}

Morphological gas properties were discussed by \cite{Owers2014,Kempner2003}. Fig.~\ref{fig:owers.jpg} (top panel) shows the X-ray emission from \textit{Chandra}, highlighting the gas morphology: peak and edge and their positions in the plane of the sky. The bottom left panel of Fig.~\ref{fig:owers.jpg} shows the mock X-ray image produced from the simulation, along with contours of total mass. In bottom right panel, we show the emission-weighted temperature map exhibiting the pronounced shock fronts, and total mass contours from the simulation.
 
In order to produce mock X-ray surface brightness maps, we use the python package pyXSIM \citep{zuhone2014,2012Biffi1,biffi2013}.
These mock images allows us to make a comparison between the simulated model and the observed X-ray configuration of A2034.
The pyXSIM works in three stages: generation of photons, projection of the generated photons and detector response to the created event.
The input to generate the photons comprises the simulation file  and the physical properties, such as cluster coordinates (RA = 15$^{\rm h}$10$^{\rm m}$11$^{\rm s}$\!.93, Dec.= 33$^{\circ}$30$^\prime$36$^\prime{^\prime}$\!.77), redshift ($z$ = 0.114) and abundance (0.29 $Z_{\odot}$). The number of photons to be generated is controlled from the specified exposure time, collecting area, and redshift. The second stage consists of projecting the photons created in the line of sight. For this it is necessary to specify the hydrogen column to be considered for the foreground Galactic absorption. Following \cite{Owers2014}, we use $N_H$ = $1.58\times10^{20} {\rm cm}^{-2}$. After creating and projecting the photons along the line of sight, the next step is to enter the parameters to obtain the response from the detector. In our case, we employ \textit{Chandra} ACIS-I, with 250 ks of exposure time and 0.5--7.0 keV in energy band. The output is a FITS file that can be processed with standard software tools for working with X-ray observations. Observational and detector properties were obtained from \cite{Owers2014} to generate the mock of A2034 simulation. The details of procedure for emulating X-ray sources with pyXSIM can be found in \cite{zuhone2014}.

Different simulation parameters generate different gas morphologies in X-ray mocks. The bottom panels in Fig.~\ref{fig:owers.jpg} show the simulated gas properties (morphology and temperature) for the A2034 at the moment when constraints of separation between mass peaks were reached.
We can note a similar simulated morphology (center panel), where emission peak in simulated X-rays roughly coincides with the position of the observed peak. In the observations \citep{Owers2014} the temperature of the shock front is around 10--12 keV, depending on which sector is considered, and drops to about 5 keV but the error bars are considerable. In our model the temperature shows a sharp discontinuity in same region, dropping from $\sim$12 keV to $\sim$2 keV.

In the next section, we will discuss the dissociative effects of this collision scenario, without the specific goal of modelling A2034 in particular. The theoretical discussion is based on different models from different scale lengths of gas and DM, and their correlations with the initial central densities.

\begin{figure*}
	\includegraphics[width=\textwidth]{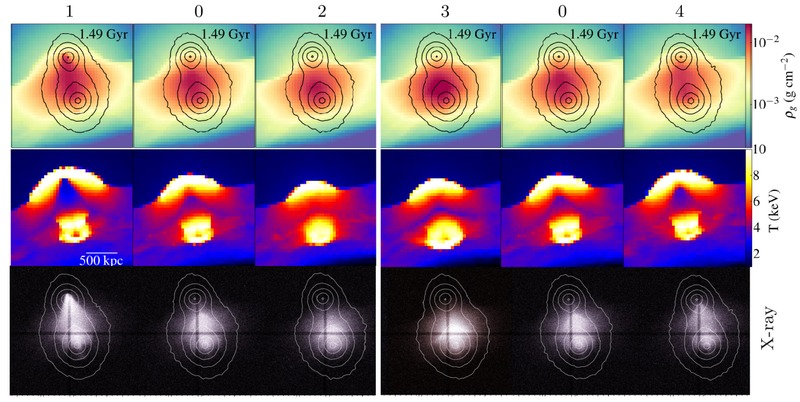}
    \caption{Gas scale length variations (from values in  Table~\ref{tab:tab_scale}), for north and south substructures. The first row shows gas density maps with contours representing total mass. The second row shows projected temperature maps for each model. The third row shows X-rays mocks from counts/pixel and contour curves of the total mass.}
    \label{fig:a_gas.jpg}
\end{figure*}

\begin{figure*}
	\includegraphics[width=\textwidth]{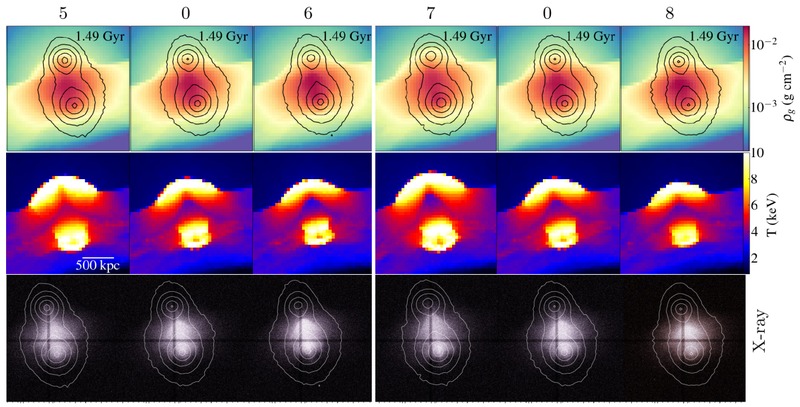}
    \caption{Dark matter halo scale length variations for north and south substructure. Similar to Fig.~\ref{fig:a_gas.jpg}, the contour curves represent total mass in the gas X-ray mock and projected density maps.}
    \label{fig:a_halo_.jpg}
\end{figure*}

\subsection{What determines the level of dissociation}
\label{sec:dissociative}

Having obtained a model that reproduces the observed properties of the A2034 cluster merger, we will now expand the discussion on the dissociative features of the collision, investigating the effects of different scale lengths in the degree of dissociation. In particular, we will measure the DM and gas separation and also the relative retention of gas within each cluster. The purpose of this section is not to perform a fine-tuning of the best model for A2034. Rather, these variations are meant to be a general theoretical exploration of the parameters that may influence the level of dissociation. Nevertheless, for the sake of consistency in nomenclature, we will continue to refer to the simulated clusters as `the north' and `the south', keeping the same orientation of the substructures. The south is the main cluster, while the north is the `bullet', so to speak.

The present discussion is based on a theoretical exploration on the relationship between the initial central densities of gas and dark matter with different dissociation levels. Therefore, exploring different scale lengths of gas and DM, maintaining the other fixed parameters ($b$, $v$, $i$ and mass ratio), we are investigating the effects that different scale lengths cause on the collision. We will refer qualitatively to models with small scale lengths as being `more concentrated', and to models with large scale lengths as being `less concentrated'. These expressions do not refer to the parameter $c$ specifically, and they apply separately for the dark matter and for the gas.

We vary the scale length of the gas and dark matter halo for the two substructures, around the default model `0'. The default scale length is defined according to with the expected concentration $c$, given the mass and redshift \citep{Duffy2008} of each substructure. Expected northern scale length is $a_{\rm g}$ =  $a_{\rm h}$ = 300 kpc and the southern scale length is $a_{\rm g}$ =  $a_{\rm h}$ = 400 kpc. From the default scale length for the north and south subclusters, Table~\ref{tab:tab_scale} shows the scale length values for each model labeled 0--8, to be discussed.

Given the north and south scale length of model 0, we consider a smaller and a larger scale length around this value for gas and DM, for both subclusters. In order to investigate the effects that each scale length individually generates on dissociation, in each simulation only one parameter is varied. That is, in models 1, 0, 2 the north  gas scale length is varied, keeping the southern cluster and halo scale length fixed. Likewise, in the models 3, 0, 4 the gas scale length of the substructure south is varied, keeping the northern cluster and the DM scale length fixed at the established values. The dark matter halo scale length for north (5, 0, 6) and south (7, 0, 8) is varied in the same way, maintaining the gas scale length fixed and varying one scale length of each substructure for model.

Each model represents a different collision. Considering that each model represents a simulation of the system at 0.26 Gyr after the central passage, Fig.~\ref{fig:a_gas.jpg} presents gas properties, through density, temperature  maps and X-ray mocks for all models whose gas scale length is varied. The different gas scale length in the northern cluster (left models 1, 0, 2) and the southern cluster (right, models 3, 0, 4) generate different effects on gas morphology, and different levels of dissociation. Similarly, Fig.~\ref{fig:a_halo_.jpg} shows the effect that different DM scale lengths cause in the behavior of the gas. Each column has a scale length, representing the DM more or less concentrated, for the northern cluster (left panel, models 5, 0, 6) and southern cluster (right panel, models 7, 0, 8). 

We can notice that both the different gas (Fig.~\ref{fig:a_gas.jpg}) and DM (Fig.~\ref{fig:a_halo_.jpg}) scale lengths cause different effects on the gas in terms of density, temperature and morphology, in different dissociative levels. We will discuss in the next subsections the systematic analysis about the differences in each case, from quantitative parameters that determine a dissociative system, in terms of gas mass retention and displacement.

\begin{figure}
	\includegraphics[width=\columnwidth]{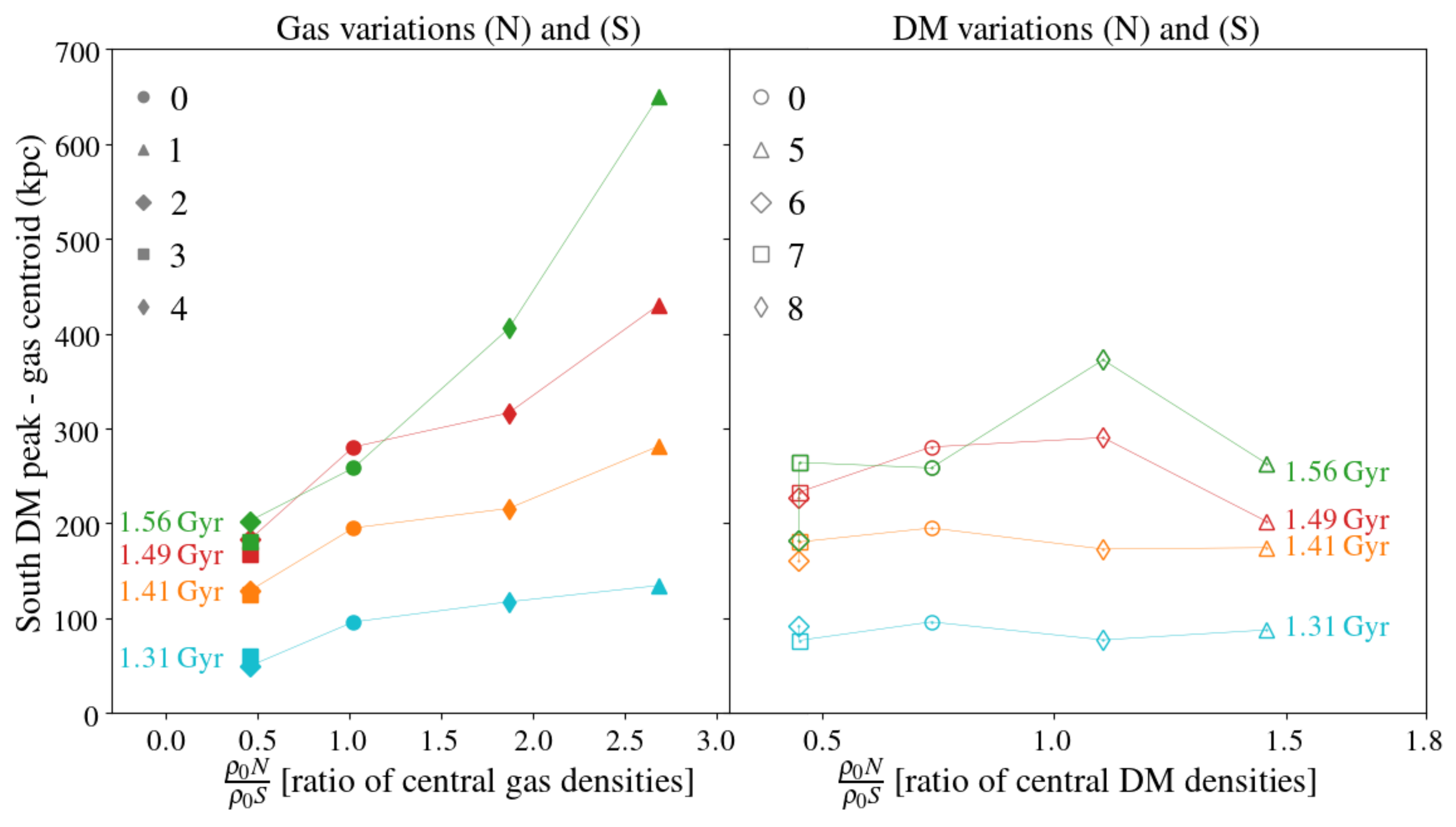}
	\caption{Gas displacement in relation to the southern dark matter peak in each model. Left: Distance of gas emission centroid to south DM  as a function of the ratio of the central gas densities. Right: Distance of gas emission centroid to south DM as a function of the ratio of the central DM densities. The icons in both panels represent the models with different gas and DM scale length and the colours represent four simulation times after the central passage.}
    \label{fig:plotdistances.jpg}
\end{figure}

\subsubsection{DM and gas separation}
One of the main features of a dissociative collision is the offset between the total mass peak and the X-ray emission peak.
To investigate the effects of the gas and dark matter scale length in the displacement of the gas mass, first
we measured the ratio of initial central densities of each model considering N/S. That is, at $t = 0$\,Gyr, we obtained the average density within a radius of 50 kpc around the centre of each substructure with different scale lengths, defined in Table~\ref{tab:tab_scale}. This gives us two important quantities for each collision: the ratio of central gas densities, and the ratio of central DM densities.

Fig.~\ref{fig:plotdistances.jpg} quantifies the displacement of gas emission in each model depending on the initial central densities of the gas (left) and DM (right). The correlation is shown for four simulation times after the central passage: two moments before and one after the best time ($t$ = 1.49 Gyr). For the purpose of measuring the gas offset in the models, we will evaluate the distance between the DM southern peak to the gas centroid in all the models which the gas and DM scale length is varied. We defined X-ray centroid as the average position of gas particles weighted by X-ray emission. 

We will analyse the two panels of Fig.~\ref{fig:plotdistances.jpg} separately, starting with the models where the gas is varied: 1, 0, 2 for the northern cluster and 3, 0, 4 for the southern cluster; both are shown in the left panel of Fig.~\ref{fig:plotdistances.jpg}. We found that the level of displacement of the gas is strongly related to the central gas density ratios. We can notice that the models that present greater displacement from the south DM to the emission centroid, are the models that have the most concentrated north gas (model 1), and the less concentrated south gas (model 4), in all times.
In contrast, the models in which the north is less concentrated and the south is more concentrated, present a greater gas retention in the south subcluster. It is notable that the displacement of gas in southern cluster is clearly associated with the gas scale length in all models. It is possible to visually verify the gas morphology in these different models in Fig.~\ref{fig:a_gas.jpg}.

Observing the X-ray mock images in Fig.~\ref{fig:a_gas.jpg}, we can see the scale length progression of the north models (left) and south models (right). The gas concentration decreases from left to right for north and south models, where the southern DM peak gradually decrease its gas centroid distance, as the north concentration decreases and the south concentration increases. This means that the greatest dissociation occurs when a high-density bullet crosses a lower-density environment (regarding gas densities). Conversely, if the clusters have comparable inner gas densities, the dissociation is less pronounced. It is worth noting that a mere factor of $\sim3$ in the ratio of central gas densities causes differences of hundreds of kpc in the offsets. This relationship is quantitatively consistent across all models, as shown in Fig.~\ref{fig:plotdistances.jpg}, where there is a strong correlation between the gas displacement depending on the concentration in the models with different gas scale lengths.

Notice that for DM scale length variations in the right panel of Fig.~\ref{fig:plotdistances.jpg}, the dependence between the mass peaks distances is much less pronounced given the different scale lengths for two substructures. That is, the gas centroid displacement is not strongly linked to the concentration of the dark matter halo, quantitatively. This indicates that the relationship between DM scale length is not as important as the gas scale length for the displacement of gas. Visually, Fig.~\ref{fig:a_halo_.jpg} shows the gas morphology in these models, where the DM concentration decrease of left to right in all frames, for north (left panel) and south (right panel) models with different DM scale lengths. The X-ray emission morphology indicates small changes from the different DM scale lengths.

It is also notable that the results hold for different times elapsed after the central passage, in all cases. In both panels of Fig.~\ref{fig:plotdistances.jpg}, the separations naturally increase with time, but the dependences remain consistent. Explorations of the distance between mass peaks (gas and DM) in all explored models indicate that the ratio of gas densities is more important than the ratio of DM densities.

\begin{figure}
	\includegraphics[width=\columnwidth]{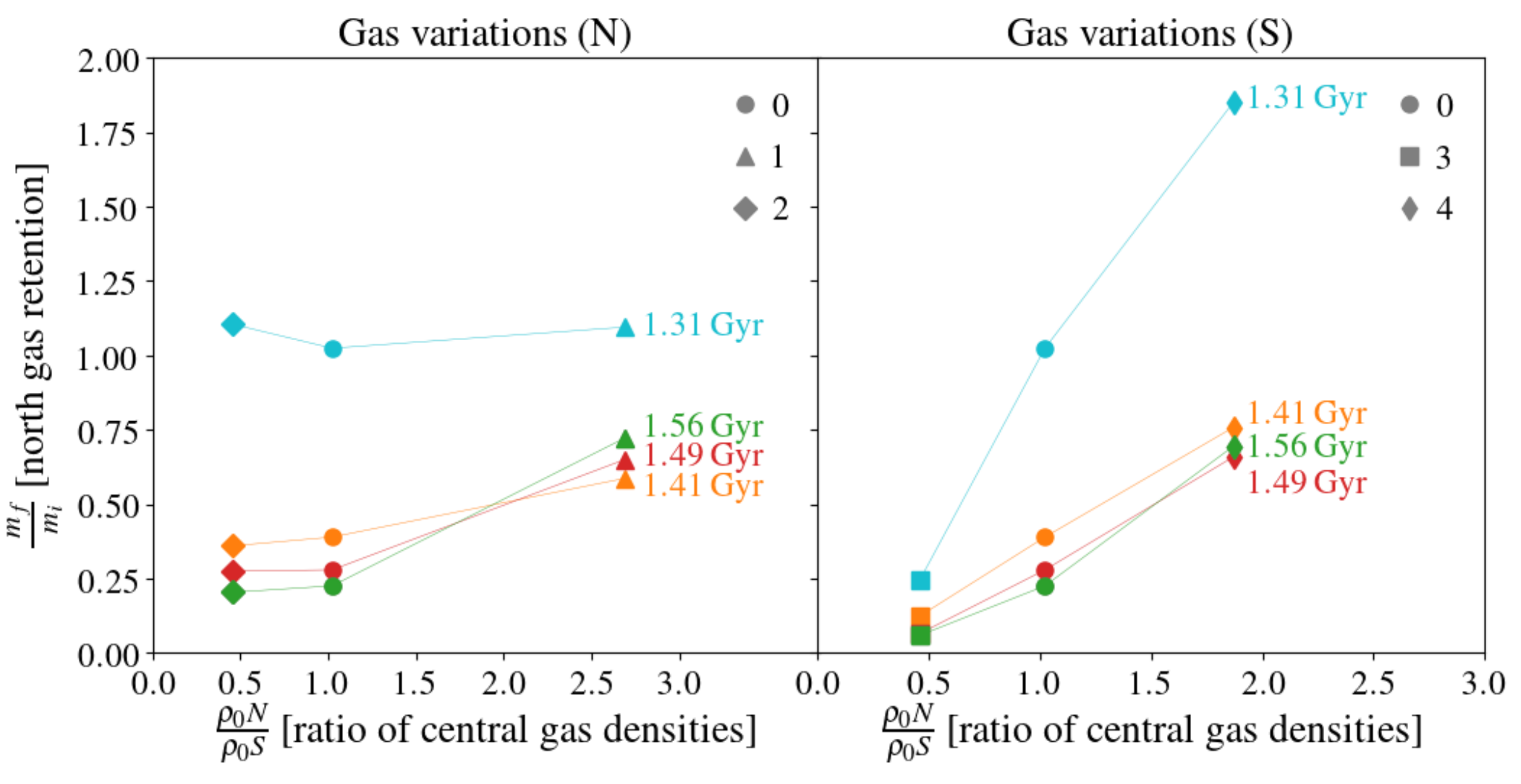}
    \caption{Mass retention of gas in the northern cluster for the models with different gas scale lengths, in four simulation times after the central passage. Left: North models with different gas scale lengths around the default. Right: South models with different gas scale lengths around the default. In both panels the gas retention depends on the ratio of central gas density.}
    \label{fig:norte_gas.jpg}
\end{figure}

\subsubsection{Gas retention: dependence on gas concentrations}

It is possible to quantify the retention of gas mass within each cluster core after the collision. Due to the interaction between the clusters, they lose more or less their gas depending on the parameters of the merger. We will quantify the loss and retention of gas in each substructure.

In order to measure the loss of gas in each substructure after the central passage, we first measured the initial mass of each cluster individually. Since the north cluster is smaller, we adopt the enclosed mass within a sphere of 100 kpc as initial gas mass for this substructure at $t$ = 0 Gyr. For the southern substructure, we employ the initial mass contained within 200 kpc. Next, we compare those with the final mass for the same radii at different times after the central passage. As we did in the previous subsections, here we also explore the relationship between gas retention and the ratios of central densities. We discuss the gas retention, first in the northern subcluster, and then, in the southern subcluster for models where the gas concentration is varied. 

Fig.~\ref{fig:norte_gas.jpg} estimates the gas remaining in the north, from the models with different north (1, 0, 2) and south (3, 0, 4) gas scale lengths. We are seeing what happens to the northern cluster given the different gas concentrations in the north (left panel) and south (right panel) after the central passage in four moments. It is important to emphasise that the gas contained in each cluster core changes over time. That is, after the central passage, the gas particles contained in 100 kpc in the northern cluster are all particles within that radius, and not necessarily those particles that originally belonged to the north at $t$ = 0 Gyr and in moments before the central passage. This is valid for both subclusters and all times after the central passage. In fact, at moments close to the central passage, when the two clusters overlap, the sphere of 100 kpc will naturally encompass more mas than at $t$ = 0 Gyr. This is the reason why $m_f/m_i$ can be greater than 1 at certain times.

\begin{figure}
	\includegraphics[width=\columnwidth]{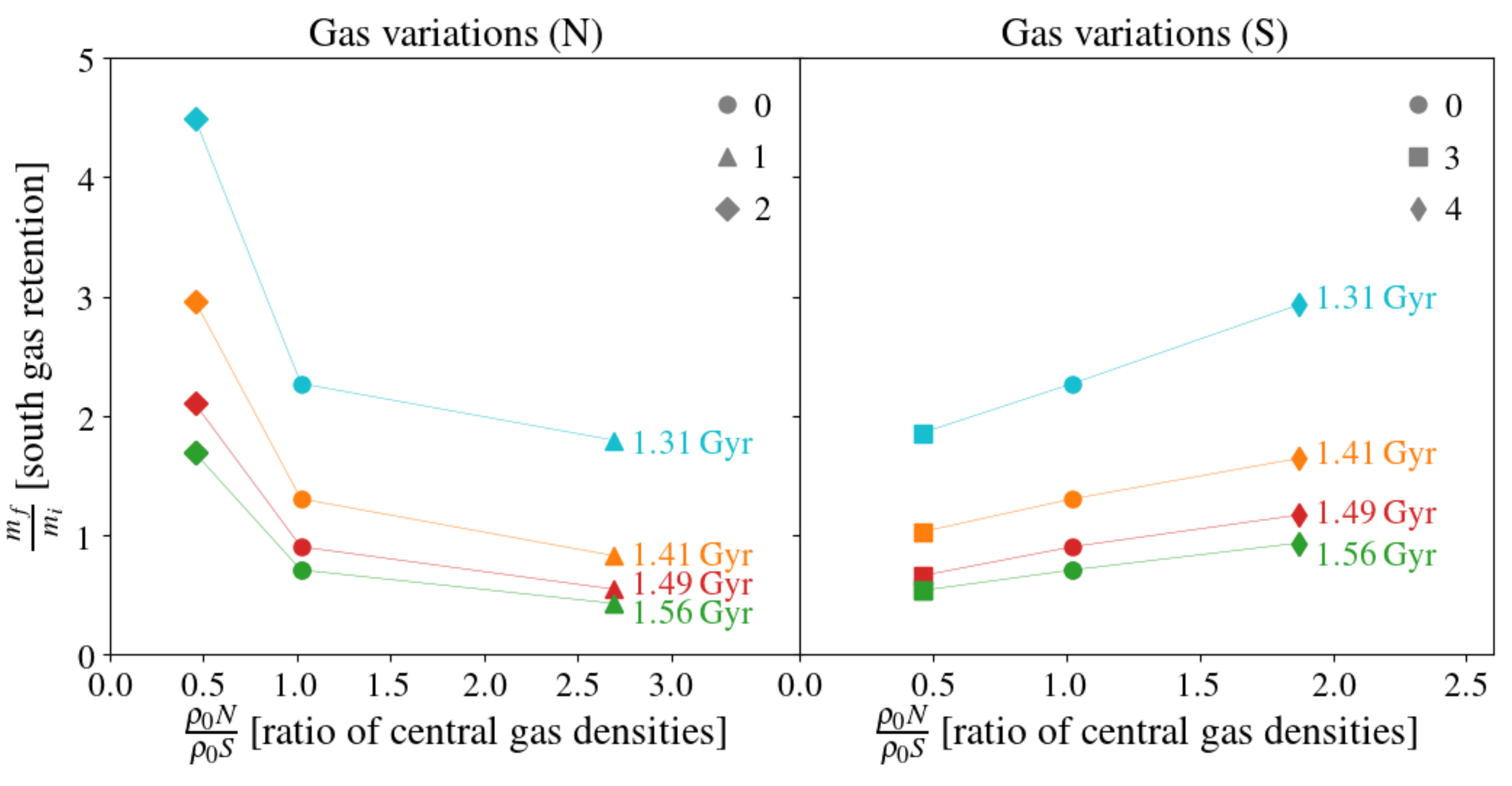}
    \caption{Mass retention of gas in the southern cluster for the models with different gas scale lengths, in four simulation times after the central passage. Left: North models with different gas scale lengths around the default. Right: South models with different gas scale lengths around the default. In both panels the gas retention depends on the ratio of central gas density.}
    \label{fig:sul_gas.pdf}
\end{figure}

About the gas retention in the north (Fig.~\ref{fig:norte_gas.jpg}), the results suggest that the more concentrated the south and less concentrated the north, the greater the probability is of the north subcluster losing its gas. This information is corroborated when we analyse the gas morphology evolution in Fig.~\ref{fig:a_gas.jpg}. In these models where the gas scale length is varied, we can notice the difference between extreme north concentration models, where model 1 represents the highest gas concentration and therefore has high retention in the north. Besides, model 2 has less concentrated gas (large scale lenght) and causes less retention in the north. At the same time, the opposite occurs for southern concentrations (right panel of Fig.~\ref{fig:a_gas.jpg}). The models with larger north scale length and smaller south scale length, cause greater gas loss in the north subcluster in all moments explored after the central passage. In other words, a more concentrated gas bullet is able to retain most of its gas, whereas a low-density bullet will lose most of its gas when passing through a dense medium.

It is also possible to notice that the amount of gas present in the models changes over time. The amount of gas in the different models at $t$ = 1.31 Gyr is greater than the amount of gas in the other times after this. Bearing in mind that the central passage occurs at $t$ = 1.23 Gyr, the next time evaluated was only 0.08 Gyr after the core passage. Over time, the amount of gas has stabilised as the clusters move away, showing a greater regularity, as we see in time progression in Fig.~\ref{fig:norte_gas.jpg}.

Regarding the gas retention in the south subcluster, similar to the north retention, the question is what happens with gas in the south cluster, given the different gas scale lengths in the north and south at different times of simulation. Notice that in Fig.~\ref{fig:sul_gas.pdf} the $m_f/m_i$ are often greater than 1, meaning that the south now holds additional gas that was abandoned by the north. Therefore this quantity also measures the accretion of gas in the south, rather than retention in such cases. Nevertheless, it is possible to compare the final relative gas content among models.

First we evaluate the gas content in the southern cluster on the left panel of Fig.~\ref{fig:sul_gas.pdf}. The configuration indicates how the gas content in the southern subcluster is related to the scale lengths of the northern cluster. It is possible to notice that the model with the most concentrated north gas (model 1) holds less gas in the south cluster. In other words, the model with the smallest north gas scale length causes greater dissociation in the south cluster. Besides, the north model with the least concentrated gas (model 2, larger scale length) causes greater gas retention in the south subcluster. Since $m_f/m_i > 1$ for model 2, this means that the inner south cluster not only retained its original gas, but also gained additional gas which was left behind by the north cluster. The configuration suggests a regular correlation between the northern gas concentration, and the final gas content is the south. The complementarity between gas loss in the north and gain in the south is not immediately obvious, however. We are specifically measuring the gas contents in the cores (within 100 and 200 kpc respectively), while substantial portions of the gas -- the densest parts, even -- are outside of those regions. Given the complex morphology that follows the central passage, it is not evident precisely how the gaseous cores would be affected, depending on the concentration of each model. In the left panel of the Fig.~\ref{fig:a_gas.jpg}, the progress is visually noticeable: looking at the south subcluster (on the right in each frame), it increases the amount of gas as the gas scale length of the north subcluster also increases. Thus, if the bullet gas is sufficiently denser than the southern gas it crosses, it can effectively strip the south of some of its gas.

Now, we will evaluate the right panel of Fig.~\ref{fig:sul_gas.pdf}. This panel indicates the effect that different south gas scale lengths cause in the gas retention of southern cluster. As well as for different gas concentrations in the north, different gas concentrations in the south affect gas retention in the south subcluster. The model with the most concentrated south (model 3) causes less gas retention, while the model with less concentrated gas (model 4) causes greater gas retention. This relationship is not clearly noticeable when we verifying Fig.~\ref{fig:a_gas.jpg} (right panel). The southern cluster progressively increases the amount of gas within 200 kpc as the gas concentration increases (from right to left, in right panel of Fig.~\ref{fig:a_gas.jpg}). In this comparison, the south models that started with the lowest concentration seemed to be the ones who held relatively more gas in the end. This is understandable considering the relative gains refer to the initial configuration. As Fig.~\ref{fig:a_gas.jpg} indicates, the gas in the core of model 4 is in fact not denser than the other models at the end of the simulation.

\subsubsection{Gas retention: dependence on DM concentrations}

\begin{figure}
	\includegraphics[width=\columnwidth]{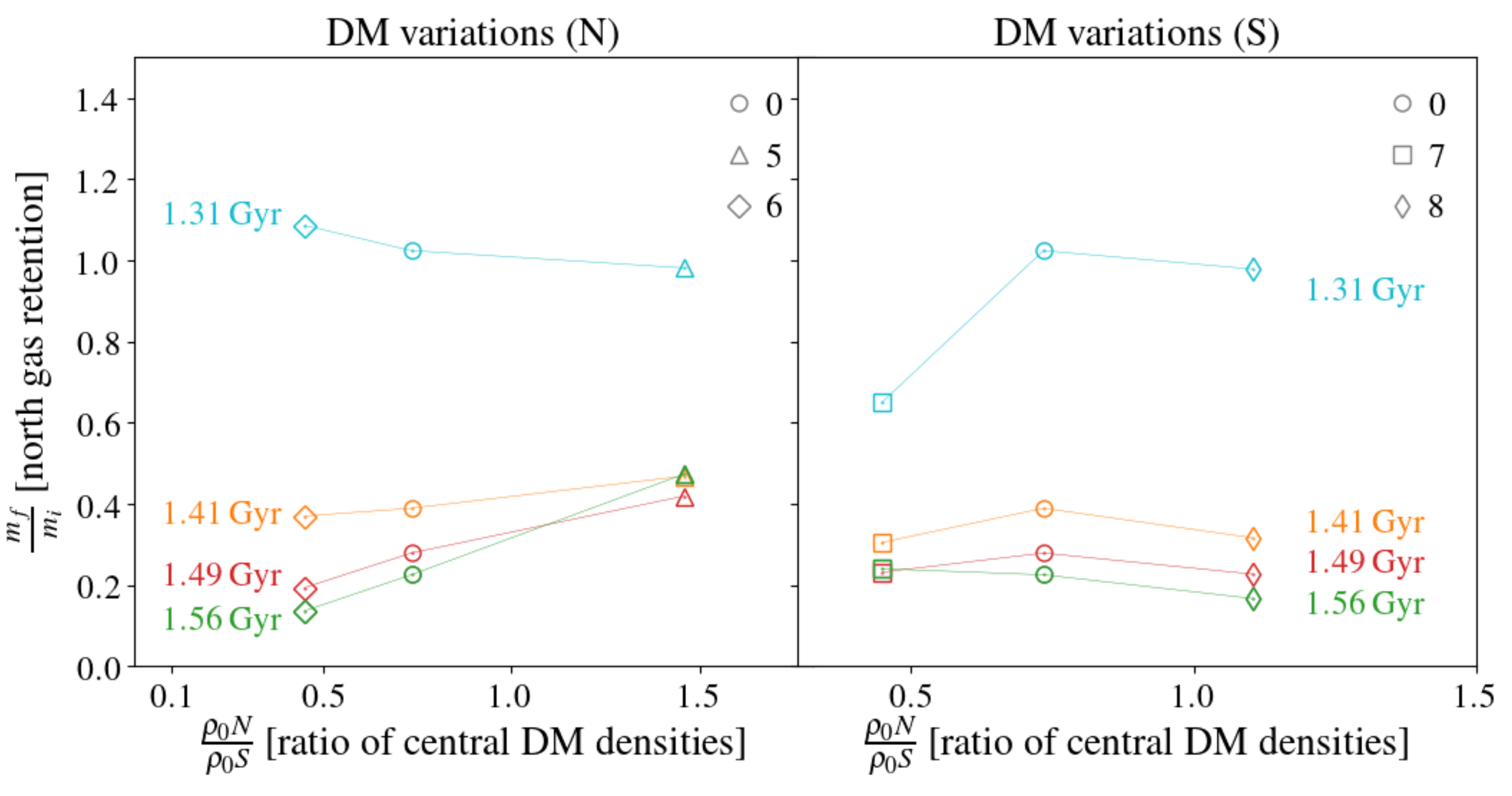}
	\caption{Mass retention of gas in the northern cluster for the models with different DM scale lengths, in four simulation times after the central passage. Left: Models with different north DM scale lengths around the model default. Right: Models with different south DM scale lengths around the model default. In both panels the gas retention depends on the ratio of central DM density.}
    \label{fig:halonorte.jpg}
\end{figure}

\begin{figure}
	\includegraphics[width=\columnwidth]{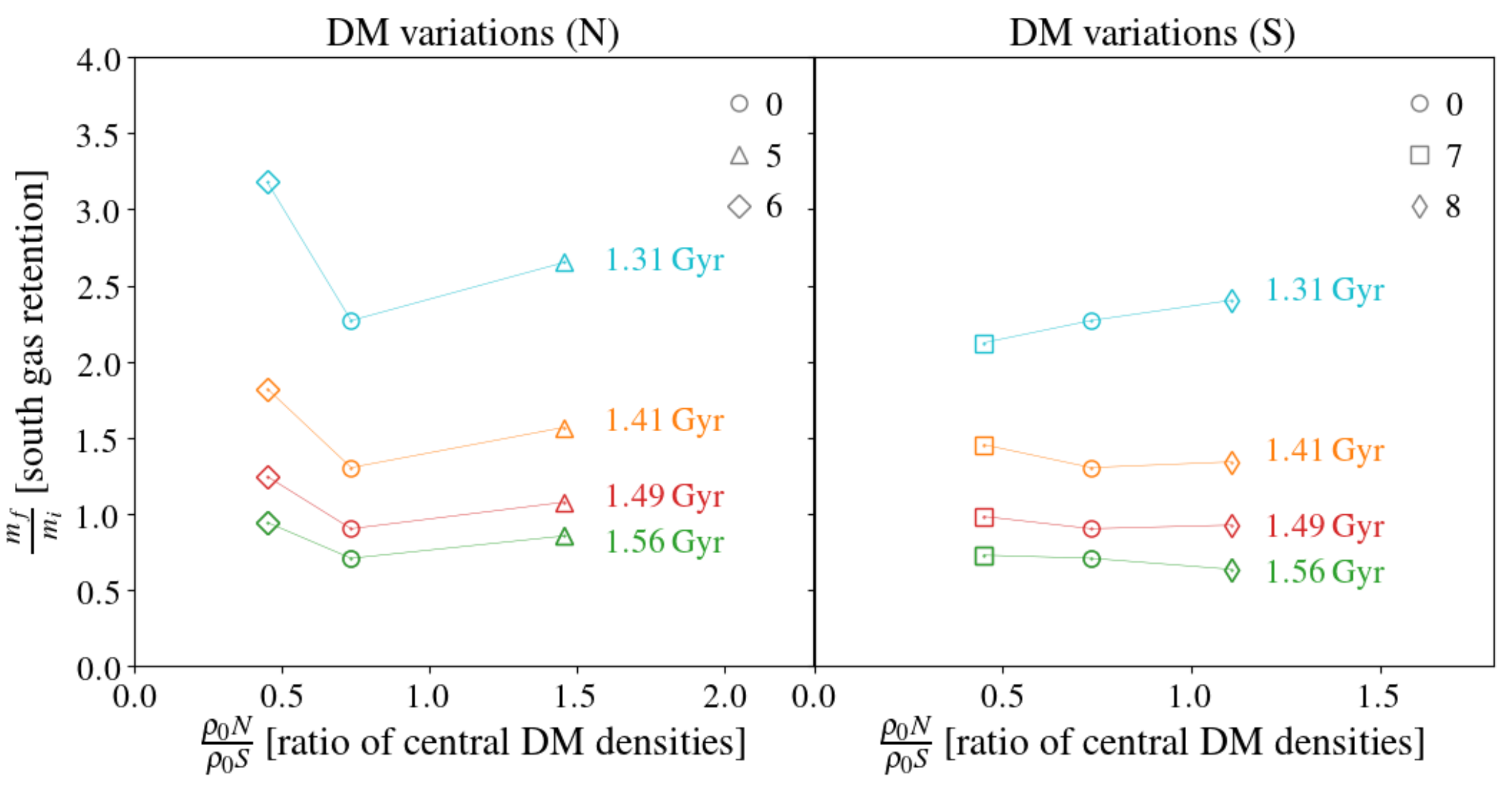}
    \caption{Mass retention of gas in the southern cluster for the models with different DM scale lengths for north (left) and south (right) subclusters. Similar to the Fig.~\ref{fig:halonorte.jpg}, in both panels the gas retention depending on the ratio of central DM densities, in four simulation times after the central passage.}
    \label{fig:sulhalo.jpg}
\end{figure}

After exploring the gas scale lengths for the two subclusters, we will discuss effects of gas retention for models where
the dark matter scale length is varied in the two subclusters. Similarly to the gas, we measure the initial and final gas mass in both clusters in order to explore their relationship with the ratio of central DM densities and their influence on dissociation. In these models with different DM scale lengths, the question is: what happens to gas in the subclusters, after the central passage given the DM scale length variations in south and north. 

For the retention of northern gas, Fig.~\ref{fig:halonorte.jpg} shows how the gas behaves from different values of north DM scale length (left) and south DM (right). Similar to the previous cases, we evaluated the gas retention for different times after the central passage. The results indicate that the DM north scale length and north gas retention are correlated. However, the dependence is less steep than the equivalent correlation with gas densities. More concentrated north DM generate more gas retention in the north subcluster, left panel of Fig.~\ref{fig:halonorte.jpg}. At the same time, the model with the least concentrated north  generates less gas retention in the north substructure. This result can be consulted visually in Fig.~\ref{fig:a_halo_.jpg}, where model 5 presents a gas mass coinciding with the north dark matter peak, that is not visually perceptible in the less concentrated model 6. Nevertheless, in the models where the southern DM in varied (Fig.~\ref{fig:halonorte.jpg}, right panel), there is no clear correlation. That is, the gas retention in the north increases with the DM concentration, but does not depend on the south DM concentration.

Concerning to the gas retention in south substructure, Fig.~\ref{fig:sulhalo.jpg} (left and right) presents different gas retention given different DM scale lengths. Analyzing the left panel, there is no clear correlation between north DM scale length dependence with south gas retention. The morphology of the gas emission in the southern cluster from different north DM scale length is shown in Fig.~\ref{fig:a_halo_.jpg} (left). The configuration suggest that regardless of variations in the northern halo, the southern gas is not significantly lost. Note that for gas retention in both subclusters, the correlations (or lack thereof) are not very time-dependent. On moments shortly after the central passage there is considerable overlap, due to the mixture of gas particles, until it stabilises in the next times. The results suggest that the gas retention in the southern cluster is not strongly related to the different DM concentrations. Analysis of gas retention for different DM scale lengths indicates that DM scale lengths are less decisive for dissociation, than gas scale lengths.

\section{Discussion}
\label{sec:conclusions}
A2034 has been interpreted as the outcome of a collision between two substructures: a more massive southern cluster and a smaller northern cluster. The merger scenario is discussed recently in \cite{Owers2014, MonteiroOliveira2018, Golovich2019}, where the configuration suggests a collision near to the plane of sky between the north and south substructures. After the core passage (with small impact parameter), the north substructure loses its gas considerably. At this moment, the observed distance between the total mass subclusters is approximately 720 kpc and the gas peak emission is offcentred in relation to the south DM peak by approximately 350\,kpc \citep{MonteiroOliveira2018}. At the same time, the X-ray observations of the system show a temperature discontinuity close to the northern BCG, that approximately coincides with the northern DM peak in this region \citep{Owers2014}. A2034 is thus a prototypical example of a dissociative cluster, in which the gas has been clearly detached from the DM, with the peak of X-ray emission lying roughly midway between the DM peaks. For these reasons, A2034 serves as a convenient reference cluster to investigate the details of the dissociation in general.

In this paper, we performed a set of hydrodynamical $N$-body simulations to study the collision of A2034. The masses of the initial conditions are based on recent gravitational weak lensing results \citep{MonteiroOliveira2018} and correspond to a mass ratio of about 1:2.2. The default models are created with the expected concentrations for those masses \citep{Duffy2008}. We presented a specific model for the collision of A2034 and we also investigated the effects that different gas and DM concentrations have on the dissociation of this system. For the first part, we did a parameter space exploration aiming to reproduce the observational constraints. For the second part, we fixed some of the parameters of the best model of A2034 ($b$, $v_0$, $i$ and mass ratio) to explore solely the effects that different concentrations (i.e. scale lengths of both gas and DM) cause on dissociation to that system.

The time after the central passage in the best model (0.26 Gyr) roughly agrees with the proposed estimates of $0.56^{+0.15}_{-0.22}$ Gyr by \cite{MonteiroOliveira2018} and of 0.3 Gyr by \cite{Owers2014}. At the time of appropriate separation, the simulated north substructure exhibits a shock front that roughly coincides with the observed edge described in \cite{Owers2014}. Given the various observed constraints, such as the appropriate morphology in the observed separation, in our model the gas peak emission it is separated from the south mass peak at $\sim280$ kpc, which agrees with the interval measured by \cite{MonteiroOliveira2018}. In this context, we presented one plausible model to reproduce the dynamic history of the dissociative collision of A2034. In this type of modelling, it is always the case that the so-called best model cannot be guaranteed to be a unique solution. For example, a similar model fine-tuned to non-zero inclination might have produced marginally improved comparisons. For simplicity, we chose to keep the $i$ = 0$^\circ$ as the default, as this also helps to disentangle the different effects when performing the second part of the analysis.

As a substructure moves through the ICM of the main cluster, it may be partially stripped of its gas due to ram pressure, specially if it moves sufficiently fast and close to the core. The diffuse gas may become unbound from its original halo, but remaining bound to the total potential of the two clusters \citep{2001Ricker}. Aiming to investigate the effects that different gas and DM concentrations cause in the level of dissociation, we explored different dissociation regimes based on the ratio of central gas and halo densities. We approached nine simulation models with different scale lengths. In each model, only one parameter was varied.
Measuring the ratio of central gas densities and the ratio of central DM densities of each pair (considering N/S) of each model at $t$ = 0 Gyr it is possible to quantify the relationship of these ratios with the dissociation. These ratios of central densities may be interpreted as a type of density contrast between the bullet and its surrounding environment, as it crosses the core of the main cluster. For instance, a large value of density ratio means a dense bullet passing through a relatively rarefied medium. This applies separately for the gas and for the DM. Indeed, it is possible to notice in terms of density and temperature maps and with X-rays mocks that different concentrations cause different effects on the gas properties after the collision. We found a diversified range of dissociation levels, where we quantify it in terms of: gas retention in both substructures and distance between the X-ray emission centroid to the DM south. All correlations were explored for four simulation times after the core passage. The strongest correlation we found was between the ratio of central gas densities and the dissociation. This relationship is clear when we observed model~1~(Fig.~\ref{fig:a_gas.jpg}), where the cometary morphology was caused by a greater gas concentration in the northern structure. It is worth noting that a difference of only 100 kpc in gas the scale length, with respect to model 0, was sufficient to drastically change the outcome. Interestingly, the same kind of behavior is observed in the mock X-ray of model 4, where a less pronounced by analogous cometary morphology also develops. In model 4 only the gas concentration in the southern cluster is varied. This is a further indication that the ratio between central gas densities is a very important factor governing the fate of the dissociation. This parameter is strongly correlated with the quantitative measure of the offset between X-ray centroid and DM peaks.

Another indicator of the level of dissociation is the stripping of gas from the cluster cores. Quantifying the relative gas retention in the two substructures in relation to central gas and DM densities, we found that gas retention in both substructures are strongly related to the gas concentrations adopted. This means that the retention of gas in the cluster cores after the collision depends sensitively on the contrast of gas density between the two clusters. On the other hand, the ratios of DM central densities showed only mild correlations or none at all. In other words, the gas retention in each substructure is not very dependent on the halo scale length adopted. Visually, Fig.~\ref{fig:a_halo_.jpg} shows that from different halo concentrations, the X-ray morphology does not change significantly for all models.

The importance of the gas density contrast can be understood in terms of the bullet's abillity to retain its cohesion if it is sufficiently denser than its surroundings. It might have been expected that the ratios of DM central densities would also have had a significant role in determining whether the gas was stripped of retained. This was not the case. The correlations between gas retention and DM central densities were mild at best. We found that if the DM potential well of the bullet is relatively shallow, its gas tends to be stripped more easily, while a deeper DM potential tends to retain more gas. But this effect is small when compared to the effects of gas densities. In the case of the main cluster, the shape of its inner DM potential well is even less relevant in determining gas loss -- at least for the ranges we explored. Naturally, the DM mass does dominate the global dynamics of the system. In our numerous variations, we altered concentrations, but not total masses. Thus, the bulk motion of the DM haloes is hardly affected.

While we systematically varied both concentrations (gas and DM) of both cluster (north and south) in all possible combinations, some properties were kept constant throughout, namely: the mass ratio, the initial velocity, the impact parameter, the baryon fraction and the inclination. The concentration dependencies we presented are thus based upon the specific model for A2034. Further exploration involving different mass ratios and other parameters is necessary to investigate more deeply the relationship between gas density ratios and the dissociative scenario in general.

Another caveat is that a third structure, A2034W, was identified as a galaxy clump by \cite{MonteiroOliveira2018} but it was not taken into account in the simulations. Dedicated simulations of cluster mergers involving three or more structures simultaneously are possible but still rare \citep{Bruggen2012, Ruggiero2019, 2020_Lia1644}. However, the mass of A2034W is not conclusively determined and its presence does not seem to be affecting the X-ray morphology of the merger. Yet the origin of the complex radio emission in that region remains unclear and may indicate more complicated shocks \citep{2016Shimwell}.

Regarding the dark matter nature and its imprints in post-merger galaxy clusters, our results are able to describe within the $\Lambda$CDM model the observed (or not) offsets in respect to the gas. However, during a cluster collision, the dark matter particles behave more closely to the non-interacting galaxies. It is relevant to measure a possible detachments between them in the same way as did for the gas \citep[e.g.][]{Ng2017} in future simulation including self-interacting dark matter models.

\section{Summary}
\label{sec:summary}

We briefly summarise our main findings regarding A2034 in particular and the level of dissociation in general:

(i) Our proposed model consists of a frontal collision between A2034S and A2034N, where the northern cluster collides with the southern substructure with a small impact parameter and close to plane of the sky (we adopt $b$ = 0 kpc, $i$ = 0$^\circ$ and $v_0$ = 2000 km/s). According to our best model, the system is being observed at 0.26 Gyr after the central passage, when the distance between the total masses peaks reaches 720 kpc.

(ii) We performed an additional set of simulations systematically varying gas and DM concentrations of both clusters. Measuring the distances between the mass peaks and the X-rays peaks, we found that these separations correlate strongly with the ratios of central gas density, but not as strongly with the ratios of central DM density. Likewise, we found that the gas retention in both clusters is not very sensitive to DM concentrations.

In conclusion, we presented a specific model that satisfactorily reproduces several observed features of A2034. We found that the ratio of central gas densities is more important than the ratio of central DM densities in determining the level of dissociation. A broad and systematic exploration of the parameter space is needed to evaluate the effects of concentrations in other regimes of dissociation. Such approach would also help address the question of under what circumstances will dissociation take place and to what degree.

\section*{Acknowledgements}
The authors acknowledge the National Laboratory for Scientific Computing (LNCC/MCTI, Brazil) for providing HPC resources of the SDumont supercomputer, which have contributed to the research results reported within this paper. This work made use of the computing facilities of the Laboratory of Astroinformatics (IAG/USP, NAT/Unicsul), whose purchase was made possible by the Brazilian agency FAPESP (grant 2009/54006-4) and the INCT-A. Also, this work made use of the computing facilities at the Centro de Computa\c c\~ao Cient\'ifica e Tecnol\'ogica (UTFPR). MTM acknowledges support from PPGFA/UTFPR. REGM acknowledges support from the Brazilian agency Conselho Nacional de Desenvolvimento Cient\'ifico e Tecnol\'ogico (CNPq) through grants 303426/2018-7 and 406908/2018-4.

\section*{Data availability}
The data underlying this article will be shared on reasonable request to the corresponding author.

\bibliographystyle{mnras}
\bibliography{A2034}

\begin{thebibliography}{}
\makeatletter
\relax
\def\mn@urlcharsother{\let\do\@makeother \do\$\do\&\do\#\do\^\do\_\do\%\do\~}
\def\mn@doi{\begingroup\mn@urlcharsother \@ifnextchar [ {\mn@doi@}
  {\mn@doi@[]}}
\def\mn@doi@[#1]#2{\def\@tempa{#1}\ifx\@tempa\@empty \href
  {http://dx.doi.org/#2} {doi:#2}\else \href {http://dx.doi.org/#2} {#1}\fi
  \endgroup}
\def\mn@eprint#1#2{\mn@eprint@#1:#2::\@nil}
\def\mn@eprint@arXiv#1{\href {http://arxiv.org/abs/#1} {{\tt arXiv:#1}}}
\def\mn@eprint@dblp#1{\href {http://dblp.uni-trier.de/rec/bibtex/#1.xml}
  {dblp:#1}}
\def\mn@eprint@#1:#2:#3:#4\@nil{\def\@tempa {#1}\def\@tempb {#2}\def\@tempc
  {#3}\ifx \@tempc \@empty \let \@tempc \@tempb \let \@tempb \@tempa \fi \ifx
  \@tempb \@empty \def\@tempb {arXiv}\fi \@ifundefined
  {mn@eprint@\@tempb}{\@tempb:\@tempc}{\expandafter \expandafter \csname
  mn@eprint@\@tempb\endcsname \expandafter{\@tempc}}}

\bibitem[\protect\citeauthoryear{{Bauer} et~al.,}{{Bauer}
  et~al.}{2015}]{Bauer2015}
{Bauer} D.,  et~al., 2015, \mn@doi [Physics of the Dark Universe]
  {10.1016/j.dark.2015.04.001}, \href
  {https://ui.adsabs.harvard.edu/abs/2015PDU.....7...16B} {7, 16}

\bibitem[\protect\citeauthoryear{{Biffi}, {Dolag}, {B{\"o}hringer}  \&
  {Lemson}}{{Biffi} et~al.}{2012}]{2012Biffi1}
{Biffi} V.,  {Dolag} K.,  {B{\"o}hringer} H.,   {Lemson} G.,  2012, \mn@doi
  [\mnras] {10.1111/j.1365-2966.2011.20278.x}, \href
  {https://ui.adsabs.harvard.edu/abs/2012MNRAS.420.3545B} {420, 3545}

\bibitem[\protect\citeauthoryear{{Biffi}, {Dolag}  \& {B{\"o}hringer}}{{Biffi}
  et~al.}{2013}]{biffi2013}
{Biffi} V.,  {Dolag} K.,   {B{\"o}hringer} H.,  2013, \mn@doi [\mnras]
  {10.1093/mnras/sts120}, \href
  {https://ui.adsabs.harvard.edu/abs/2013MNRAS.428.1395B} {428, 1395}

\bibitem[\protect\citeauthoryear{{Brada{\v c}}, {Allen}, {Treu}, {Ebeling},
  {Massey}, {Morris}, {von der Linden}  \& {Applegate}}{{Brada{\v c}}
  et~al.}{2008}]{Brada2008}
{Brada{\v c}} M.,  {Allen} S.~W.,  {Treu} T.,  {Ebeling} H.,  {Massey} R.,
  {Morris} R.~G.,  {von der Linden} A.,   {Applegate} D.,  2008, \mn@doi [\apj]
  {10.1086/591246}, \href
  {https://ui.adsabs.harvard.edu/abs/2008ApJ...687..959B} {687, 959}

\bibitem[\protect\citeauthoryear{{Breuer}, {Werner}, {Mernier}, {Mroczkowski},
  {Simionescu}, {Clarke}, {ZuHone}  \& {Di Mascolo}}{{Breuer}
  et~al.}{2020}]{2020Breuer2256}
{Breuer} J.~P.,  {Werner} N.,  {Mernier} F.,  {Mroczkowski} T.,  {Simionescu}
  A.,  {Clarke} T.~E.,  {ZuHone} J.~A.,   {Di Mascolo} L.,  2020, \mn@doi
  [\mnras] {10.1093/mnras/staa1492}, \href
  {https://ui.adsabs.harvard.edu/abs/2020MNRAS.495.5014B} {495, 5014}

\bibitem[\protect\citeauthoryear{{Br{\"u}ggen}, {van Weeren}  \&
  {R{\"o}ttgering}}{{Br{\"u}ggen} et~al.}{2012}]{Bruggen2012}
{Br{\"u}ggen} M.,  {van Weeren} R.~J.,   {R{\"o}ttgering} H.~J.~A.,  2012,
  \mn@doi [\mnras] {10.1111/j.1745-3933.2012.01304.x}, \href
  {https://ui.adsabs.harvard.edu/abs/2012MNRAS.425L..76B} {425, L76}

\bibitem[\protect\citeauthoryear{{Cavaliere} \& {Fusco-Femiano}}{{Cavaliere} \&
  {Fusco-Femiano}}{1976}]{fusco}
{Cavaliere} A.,  {Fusco-Femiano} R.,  1976, \aap, \href
  {https://ui.adsabs.harvard.edu/abs/1976A&A....49..137C} {500, 95}

\bibitem[\protect\citeauthoryear{{Chon} \& {B{\"o}hringer}}{{Chon} \&
  {B{\"o}hringer}}{2015}]{2015Chon4067}
{Chon} G.,  {B{\"o}hringer} H.,  2015, \mn@doi [\aap]
  {10.1051/0004-6361/201425143}, \href
  {https://ui.adsabs.harvard.edu/abs/2015A&A...574A.132C} {574, A132}

\bibitem[\protect\citeauthoryear{{Clowe}, {Brada{\v{c}}}, {Gonzalez},
  {Markevitch}, {Randall}, {Jones}  \& {Zaritsky}}{{Clowe}
  et~al.}{2006}]{CloweBradac2006}
{Clowe} D.,  {Brada{\v{c}}} M.,  {Gonzalez} A.~H.,  {Markevitch} M.,  {Randall}
  S.~W.,  {Jones} C.,   {Zaritsky} D.,  2006, \mn@doi [\apjl] {10.1086/508162},
  \href {https://ui.adsabs.harvard.edu/abs/2006ApJ...648L.109C} {648, L109}

\bibitem[\protect\citeauthoryear{{Dahle} et~al.,}{{Dahle}
  et~al.}{2013}]{2013ADahle}
{Dahle} H.,  et~al., 2013, \mn@doi [\apj] {10.1088/0004-637X/772/1/23}, \href
  {https://ui.adsabs.harvard.edu/abs/2013ApJ...772...23D} {772, 23}

\bibitem[\protect\citeauthoryear{{David}, {Forman}  \& {Jones}}{{David}
  et~al.}{1999}]{David1999}
{David} L.~P.,  {Forman} W.,   {Jones} C.,  1999, \mn@doi [\apj]
  {10.1086/307388}, \href
  {https://ui.adsabs.harvard.edu/abs/1999ApJ...519..533D} {519, 533}

\bibitem[\protect\citeauthoryear{{Dawson} et~al.,}{{Dawson}
  et~al.}{2012}]{Dawson2012}
{Dawson} W.~A.,  et~al., 2012, \mn@doi [\apjl] {10.1088/2041-8205/747/2/L42},
  \href {http://adsabs.harvard.edu/abs/2012ApJ...747L..42D} {747, L42}

\bibitem[\protect\citeauthoryear{{Dehnen}}{{Dehnen}}{1993}]{Dehnen1993}
{Dehnen} W.,  1993, \mnras, \href
  {http://adsabs.harvard.edu/abs/1993MNRAS.265..250D} {265, 250}

\bibitem[\protect\citeauthoryear{{Donnert}}{{Donnert}}{2014}]{2014Donnert}
{Donnert} J.~M.~F.,  2014, \mn@doi [\mnras] {10.1093/mnras/stt2291}, \href
  {https://ui.adsabs.harvard.edu/abs/2014MNRAS.438.1971D} {438, 1971}

\bibitem[\protect\citeauthoryear{{Donnert}, {Beck}, {Dolag}  \&
  {R{\"o}ttgering}}{{Donnert} et~al.}{2017}]{2017Donnert}
{Donnert} J.~M.~F.,  {Beck} A.~M.,  {Dolag} K.,   {R{\"o}ttgering} H.~J.~A.,
  2017, \mn@doi [\mnras] {10.1093/mnras/stx1819}, \href
  {https://ui.adsabs.harvard.edu/abs/2017MNRAS.471.4587D} {471, 4587}

\bibitem[\protect\citeauthoryear{{Doubrawa}, {Machado}, {Lagan{\'a}}, {Lima
  Neto}, {Monteiro-Oliveira}  \& {Cypriano}}{{Doubrawa}
  et~al.}{2020}]{2020_Lia1644}
{Doubrawa} L.,  {Machado} R.~E.~G.,  {Lagan{\'a}} T.~F.,  {Lima Neto} G.~B.,
  {Monteiro-Oliveira} R.,   {Cypriano} E.~S.,  2020, \mn@doi [\mnras]
  {10.1093/mnras/staa1051}, \href
  {https://ui.adsabs.harvard.edu/abs/2020MNRAS.495.2022D} {495, 2022}

\bibitem[\protect\citeauthoryear{{Drlica-Wagner} et~al.,}{{Drlica-Wagner}
  et~al.}{2019}]{Drlica-Wagner2019}
{Drlica-Wagner} A.,  et~al., 2019, arXiv e-prints, \href
  {https://ui.adsabs.harvard.edu/abs/2019arXiv190201055D} {p. arXiv:1902.01055}

\bibitem[\protect\citeauthoryear{{Duffy}, {Schaye}, {Kay}  \& {Dalla
  Vecchia}}{{Duffy} et~al.}{2008}]{Duffy2008}
{Duffy} A.~R.,  {Schaye} J.,  {Kay} S.~T.,   {Dalla Vecchia} C.,  2008, \mn@doi
  [\mnras] {10.1111/j.1745-3933.2008.00537.x}, \href
  {https://ui.adsabs.harvard.edu/abs/2008MNRAS.390L..64D} {390, L64}

\bibitem[\protect\citeauthoryear{{Gastaldello} et~al.,}{{Gastaldello}
  et~al.}{2014}]{2014GastaldelloSL2S}
{Gastaldello} F.,  et~al., 2014, \mn@doi [\mnras] {10.1093/mnrasl/slu058},
  \href {https://ui.adsabs.harvard.edu/abs/2014MNRAS.442L..76G} {442, L76}

\bibitem[\protect\citeauthoryear{{Geller}, {Diaferio}, {Rines}  \&
  {Serra}}{{Geller} et~al.}{2013}]{Geller2013}
{Geller} M.~J.,  {Diaferio} A.,  {Rines} K.~J.,   {Serra} A.~L.,  2013, \mn@doi
  [\apj] {10.1088/0004-637X/764/1/58}, \href
  {https://ui.adsabs.harvard.edu/abs/2013ApJ...764...58G} {764, 58}

\bibitem[\protect\citeauthoryear{{Golovich}, {Dawson}, {Wittman}, {Ogrean},
  {van Weeren}  \& {Bonafede}}{{Golovich} et~al.}{2016}]{2016AGovovichelgordo2}
{Golovich} N.,  {Dawson} W.~A.,  {Wittman} D.,  {Ogrean} G.,  {van Weeren} R.,
   {Bonafede} A.,  2016, \mn@doi [\apj] {10.3847/0004-637X/831/1/110}, \href
  {https://ui.adsabs.harvard.edu/abs/2016ApJ...831..110G} {831, 110}

\bibitem[\protect\citeauthoryear{{Golovich}, {van Weeren}, {Dawson}, {Jee}  \&
  {Wittman}}{{Golovich} et~al.}{2017}]{2017ApJGolovichelmagro}
{Golovich} N.,  {van Weeren} R.~J.,  {Dawson} W.~A.,  {Jee} M.~J.,   {Wittman}
  D.,  2017, \mn@doi [\apj] {10.3847/1538-4357/aa667f}, \href
  {https://ui.adsabs.harvard.edu/abs/2017ApJ...838..110G} {838, 110}

\bibitem[\protect\citeauthoryear{{Golovich} et~al.,}{{Golovich}
  et~al.}{2019}]{Golovich2019}
{Golovich} N.,  et~al., 2019, \mn@doi [\apj] {10.3847/1538-4357/ab2f90}, \href
  {https://ui.adsabs.harvard.edu/abs/2019ApJ...882...69G} {882, 69}

\bibitem[\protect\citeauthoryear{{Harvey}, {Massey}, {Kitching}, {Taylor}  \&
  {Tittley}}{{Harvey} et~al.}{2015}]{Harvey2015}
{Harvey} D.,  {Massey} R.,  {Kitching} T.,  {Taylor} A.,   {Tittley} E.,  2015,
  \mn@doi [Science] {10.1126/science.1261381}, \href
  {https://ui.adsabs.harvard.edu/abs/2015Sci...347.1462H} {347, 1462}

\bibitem[\protect\citeauthoryear{{Hayashi} \& {White}}{{Hayashi} \&
  {White}}{2006}]{2006HayashiEwhite}
{Hayashi} E.,  {White} S. D.~M.,  2006, \mn@doi [\mnras]
  {10.1111/j.1745-3933.2006.00184.x}, \href
  {https://ui.adsabs.harvard.edu/abs/2006MNRAS.370L..38H} {370, L38}

\bibitem[\protect\citeauthoryear{{Hernquist}}{{Hernquist}}{1990}]{Hernquist1990}
{Hernquist} L.,  1990, \mn@doi [\apj] {10.1086/168845}, \href
  {http://adsabs.harvard.edu/abs/1990ApJ...356..359H} {356, 359}

\bibitem[\protect\citeauthoryear{{Jee}, {Hughes}, {Menanteau}, {Sif{\'o}n},
  {Mandelbaum}, {Barrientos}, {Infante}  \& {Ng}}{{Jee} et~al.}{2014}]{Jee2014}
{Jee} M.~J.,  {Hughes} J.~P.,  {Menanteau} F.,  {Sif{\'o}n} C.,  {Mandelbaum}
  R.,  {Barrientos} L.~F.,  {Infante} L.,   {Ng} K.~Y.,  2014, \mn@doi [\apj]
  {10.1088/0004-637X/785/1/20}, \href
  {http://adsabs.harvard.edu/abs/2014ApJ...785...20J} {785, 20}

\bibitem[\protect\citeauthoryear{{Jee} et~al.,}{{Jee}
  et~al.}{2015}]{2015JeeLinguica}
{Jee} M.~J.,  et~al., 2015, \mn@doi [\apj] {10.1088/0004-637X/802/1/46}, \href
  {https://ui.adsabs.harvard.edu/abs/2015ApJ...802...46J} {802, 46}

\bibitem[\protect\citeauthoryear{{Jee}, {Dawson}, {Stroe}, {Wittman}, {van
  Weeren}, {Br{\"u}ggen}, {Brada{\v{c}}}  \& {R{\"o}ttgering}}{{Jee}
  et~al.}{2016}]{2016ApJeetoothblush}
{Jee} M.~J.,  {Dawson} W.~A.,  {Stroe} A.,  {Wittman} D.,  {van Weeren} R.~J.,
  {Br{\"u}ggen} M.,  {Brada{\v{c}}} M.,   {R{\"o}ttgering} H.,  2016, \mn@doi
  [\apj] {10.3847/0004-637X/817/2/179}, \href
  {https://ui.adsabs.harvard.edu/abs/2016ApJ...817..179J} {817, 179}

\bibitem[\protect\citeauthoryear{{Kempner} \& {Sarazin}}{{Kempner} \&
  {Sarazin}}{2001}]{2001_kempner2001A2034}
{Kempner} J.~C.,  {Sarazin} C.~L.,  2001, \mn@doi [\apj] {10.1086/319024},
  \href {https://ui.adsabs.harvard.edu/abs/2001ApJ...548..639K} {548, 639}

\bibitem[\protect\citeauthoryear{{Kempner}, {Sarazin}  \&
  {Markevitch}}{{Kempner} et~al.}{2003}]{Kempner2003}
{Kempner} J.~C.,  {Sarazin} C.~L.,   {Markevitch} M.,  2003, \mn@doi [\apj]
  {10.1086/376358}, \href
  {https://ui.adsabs.harvard.edu/abs/2003ApJ...593..291K} {593, 291}

\bibitem[\protect\citeauthoryear{{Kim}, {Peter}  \& {Wittman}}{{Kim}
  et~al.}{2017}]{2017_Kim}
{Kim} S.~Y.,  {Peter} A. H.~G.,   {Wittman} D.,  2017, \mn@doi [\mnras]
  {10.1093/mnras/stx896}, \href
  {https://ui.adsabs.harvard.edu/abs/2017MNRAS.469.1414K} {469, 1414}

\bibitem[\protect\citeauthoryear{{Lage} \& {Farrar}}{{Lage} \&
  {Farrar}}{2014}]{2014LageeFarrar}
{Lage} C.,  {Farrar} G.,  2014, \mn@doi [\apj] {10.1088/0004-637X/787/2/144},
  \href {https://ui.adsabs.harvard.edu/abs/2014ApJ...787..144L} {787, 144}

\bibitem[\protect\citeauthoryear{{Le Delliou}, {Marcondes}, {Lima Neto}  \&
  {Abdalla}}{{Le Delliou} et~al.}{2015}]{2015Ledelliou}
{Le Delliou} M.,  {Marcondes} R.~J.~F.,  {Lima Neto} G.~B.,   {Abdalla} E.,
  2015, \mn@doi [\mnras] {10.1093/mnras/stv1561}, \href
  {https://ui.adsabs.harvard.edu/abs/2015MNRAS.453....2L} {453, 2}

\bibitem[\protect\citeauthoryear{{Louren{\c{c}}o} et~al.,}{{Louren{\c{c}}o}
  et~al.}{2020}]{2399ana}
{Louren{\c{c}}o} A. C.~C.,  et~al., 2020, \mn@doi [\mnras]
  {10.1093/mnras/staa2464}, \href
  {https://ui.adsabs.harvard.edu/abs/2020MNRAS.498..835L} {498, 835}

\bibitem[\protect\citeauthoryear{{Machado} \& {Lima Neto}}{{Machado} \& {Lima
  Neto}}{2013}]{Machado2013}
{Machado} R.~E.~G.,  {Lima Neto} G.~B.,  2013, \mn@doi [\mnras]
  {10.1093/mnras/stt127}, \href
  {http://adsabs.harvard.edu/abs/2013MNRAS.430.3249M} {430, 3249}

\bibitem[\protect\citeauthoryear{{Machado} \& {Lima Neto}}{{Machado} \& {Lima
  Neto}}{2015}]{2015_2056Machado}
{Machado} R. E.~G.,  {Lima Neto} G.~B.,  2015, \mn@doi [\mnras]
  {10.1093/mnras/stu2669}, \href
  {https://ui.adsabs.harvard.edu/abs/2015MNRAS.447.2915M} {447, 2915}

\bibitem[\protect\citeauthoryear{{Machado}, {Monteiro-Oliveira}, {Lima Neto}
  \& {Cypriano}}{{Machado} et~al.}{2015}]{Machado2015b}
{Machado} R.~E.~G.,  {Monteiro-Oliveira} R.,  {Lima Neto} G.~B.,   {Cypriano}
  E.~S.,  2015, \mn@doi [\mnras] {10.1093/mnras/stv1162}, \href
  {http://adsabs.harvard.edu/abs/2015MNRAS.451.3309M} {451, 3309}

\bibitem[\protect\citeauthoryear{{Mahdavi}, {Hoekstra}, {Babul}, {Balam}  \&
  {Capak}}{{Mahdavi} et~al.}{2007}]{Mahdavi2007}
{Mahdavi} A.,  {Hoekstra} H.,  {Babul} A.,  {Balam} D.~D.,   {Capak} P.~L.,
  2007, \mn@doi [\apj] {10.1086/521383}, \href
  {http://adsabs.harvard.edu/abs/2007ApJ...668..806M} {668, 806}

\bibitem[\protect\citeauthoryear{{Markevitch} \& {Vikhlinin}}{{Markevitch} \&
  {Vikhlinin}}{2007}]{2007Markevitch}
{Markevitch} M.,  {Vikhlinin} A.,  2007, \mn@doi [\physrep]
  {10.1016/j.physrep.2007.01.001}, \href
  {https://ui.adsabs.harvard.edu/abs/2007PhR...443....1M} {443, 1}

\bibitem[\protect\citeauthoryear{{Massey}, {Kitching}  \& {Nagai}}{{Massey}
  et~al.}{2011}]{Massey2011}
{Massey} R.,  {Kitching} T.,   {Nagai} D.,  2011, \mn@doi [\mnras]
  {10.1111/j.1365-2966.2011.18246.x}, \href
  {https://ui.adsabs.harvard.edu/abs/2011MNRAS.413.1709M} {413, 1709}

\bibitem[\protect\citeauthoryear{{Mastropietro} \& {Burkert}}{{Mastropietro} \&
  {Burkert}}{2008}]{2008MastropietroeBurket}
{Mastropietro} C.,  {Burkert} A.,  2008, \mn@doi [\mnras]
  {10.1111/j.1365-2966.2008.13626.x}, \href
  {https://ui.adsabs.harvard.edu/abs/2008MNRAS.389..967M} {389, 967}

\bibitem[\protect\citeauthoryear{{Merten}, {Coe}, {Dupke}, {Massey}, {Zitrin},
  {Cypriano}  et~al.}{{Merten} et~al.}{2011}]{Merten2011}
{Merten} J.,  {Coe} D.,  {Dupke} R.,  {Massey} R.,  {Zitrin} A.,  {Cypriano}
  E.~S.,   et~al., 2011, \mn@doi [\mnras] {10.1111/j.1365-2966.2011.19266.x},
  \href {http://adsabs.harvard.edu/abs/2011MNRAS.417..333M} {417, 333}

\bibitem[\protect\citeauthoryear{{Molnar} \& {Broadhurst}}{{Molnar} \&
  {Broadhurst}}{2015}]{2015MonlareBroad}
{Molnar} S.~M.,  {Broadhurst} T.,  2015, \mn@doi [\apj]
  {10.1088/0004-637X/800/1/37}, \href
  {https://ui.adsabs.harvard.edu/abs/2015ApJ...800...37M} {800, 37}

\bibitem[\protect\citeauthoryear{{Molnar} \& {Broadhurst}}{{Molnar} \&
  {Broadhurst}}{2017}]{2017MolnareBroad}
{Molnar} S.~M.,  {Broadhurst} T.,  2017, \mn@doi [\apj]
  {10.3847/1538-4357/aa70a3}, \href
  {https://ui.adsabs.harvard.edu/abs/2017ApJ...841...46M} {841, 46}

\bibitem[\protect\citeauthoryear{{Molnar} \& {Broadhurst}}{{Molnar} \&
  {Broadhurst}}{2018}]{Molnar2018}
{Molnar} S.~M.,  {Broadhurst} T.,  2018, \mn@doi [\apj]
  {10.3847/1538-4357/aad04c}, \href
  {https://ui.adsabs.harvard.edu/abs/2018ApJ...862..112M} {862, 112}

\bibitem[\protect\citeauthoryear{{Monteiro-Oliveira}, {Cypriano}, {Machado},
  {Lima Neto}, {Ribeiro}, {Sodr{\'e}}  \& {Dupke}}{{Monteiro-Oliveira}
  et~al.}{2017}]{Monteiro2017a}
{Monteiro-Oliveira} R.,  {Cypriano} E.~S.,  {Machado} R.~E.~G.,  {Lima Neto}
  G.~B.,  {Ribeiro} A.~L.~B.,  {Sodr{\'e}} L.,   {Dupke} R.,  2017, \mn@doi
  [\mnras] {10.1093/mnras/stw3238}, \href
  {http://adsabs.harvard.edu/abs/2017MNRAS.466.2614M} {466, 2614}

\bibitem[\protect\citeauthoryear{{Monteiro-Oliveira}, {Cypriano}, {Vitorelli},
  {Ribeiro}, {Sodr{\'e}}, {Dupke}  \& {Mendes de Oliveira}}{{Monteiro-Oliveira}
  et~al.}{2018}]{MonteiroOliveira2018}
{Monteiro-Oliveira} R.,  {Cypriano} E.~S.,  {Vitorelli} A.~Z.,  {Ribeiro}
  A.~L.~B.,  {Sodr{\'e}} L.,  {Dupke} R.,   {Mendes de Oliveira} C.,  2018,
  \mn@doi [\mnras] {10.1093/mnras/sty2349}, \href
  {http://adsabs.harvard.edu/abs/2018MNRAS.481.1097M} {481, 1097}

\bibitem[\protect\citeauthoryear{{Navarro}, {Frenk}  \& {White}}{{Navarro}
  et~al.}{1997}]{1997NFW}
{Navarro} J.~F.,  {Frenk} C.~S.,   {White} S. D.~M.,  1997, \mn@doi [\apj]
  {10.1086/304888}, \href
  {https://ui.adsabs.harvard.edu/abs/1997ApJ...490..493N} {490, 493}

\bibitem[\protect\citeauthoryear{{Ng}, {Pillepich}, {Wittman}, {Dawson},
  {Hernquist}  \& {Nelson}}{{Ng} et~al.}{2017}]{Ng2017}
{Ng} K.~Y.,  {Pillepich} A.,  {Wittman} D.,  {Dawson} W.~A.,  {Hernquist} L.,
  {Nelson} D.~R.,  2017, arXiv e-prints, \href
  {https://ui.adsabs.harvard.edu/abs/2017arXiv170300010N} {p. arXiv:1703.00010}

\bibitem[\protect\citeauthoryear{{Okabe} \& {Umetsu}}{{Okabe} \&
  {Umetsu}}{2008}]{Okabe2008}
{Okabe} N.,  {Umetsu} K.,  2008, \mn@doi [Publications of the Astronomical
  Society of Japan] {10.1093/pasj/60.2.345}, \href
  {https://ui.adsabs.harvard.edu/abs/2008PASJ...60..345O} {60, 345}

\bibitem[\protect\citeauthoryear{{Okabe}, {Bourdin}, {Mazzotta}  \&
  {Maurogordato}}{{Okabe} et~al.}{2011}]{2011okabe2163}
{Okabe} N.,  {Bourdin} H.,  {Mazzotta} P.,   {Maurogordato} S.,  2011, \mn@doi
  [\apj] {10.1088/0004-637X/741/2/116}, \href
  {https://ui.adsabs.harvard.edu/abs/2011ApJ...741..116O} {741, 116}

\bibitem[\protect\citeauthoryear{{Owers} et~al.,}{{Owers}
  et~al.}{2014}]{Owers2014}
{Owers} M.~S.,  et~al., 2014, \mn@doi [\apj] {10.1088/0004-637X/780/2/163},
  \href {https://ui.adsabs.harvard.edu/abs/2014ApJ...780..163O} {780, 163}

\bibitem[\protect\citeauthoryear{{Poole}, {Fardal}, {Babul}, {McCarthy},
  {Quinn}  \& {Wadsley}}{{Poole} et~al.}{2006}]{2006PooleI}
{Poole} G.~B.,  {Fardal} M.~A.,  {Babul} A.,  {McCarthy} I.~G.,  {Quinn} T.,
  {Wadsley} J.,  2006, \mn@doi [\mnras] {10.1111/j.1365-2966.2006.10916.x},
  \href {https://ui.adsabs.harvard.edu/abs/2006MNRAS.373..881P} {373, 881}

\bibitem[\protect\citeauthoryear{{Poole}, {Babul}, {McCarthy}, {Fardal},
  {Bildfell}, {Quinn}  \& {Mahdavi}}{{Poole} et~al.}{2007}]{2007_poole2}
{Poole} G.~B.,  {Babul} A.,  {McCarthy} I.~G.,  {Fardal} M.~A.,  {Bildfell}
  C.~J.,  {Quinn} T.,   {Mahdavi} A.,  2007, \mn@doi [\mnras]
  {10.1111/j.1365-2966.2007.12107.x}, \href
  {https://ui.adsabs.harvard.edu/abs/2007MNRAS.380..437P} {380, 437}

\bibitem[\protect\citeauthoryear{{Ragozzine}, {Clowe}, {Markevitch}, {Gonzalez}
   \& {Brada{\v c}}}{{Ragozzine} et~al.}{2012}]{Ragozzine2012}
{Ragozzine} B.,  {Clowe} D.,  {Markevitch} M.,  {Gonzalez} A.~H.,   {Brada{\v
  c}} M.,  2012, \mn@doi [\apj] {10.1088/0004-637X/744/2/94}, \href
  {http://adsabs.harvard.edu/abs/2012ApJ...744...94R} {744, 94}

\bibitem[\protect\citeauthoryear{{Ricker} \& {Sarazin}}{{Ricker} \&
  {Sarazin}}{2001}]{2001Ricker}
{Ricker} P.~M.,  {Sarazin} C.~L.,  2001, \mn@doi [\apj] {10.1086/323365}, \href
  {https://ui.adsabs.harvard.edu/abs/2001ApJ...561..621R} {561, 621}

\bibitem[\protect\citeauthoryear{{Ruggiero}, {Machado}, {Roman-Oliveira},
  {Chies-Santos}, {Lima Neto}, {Doubrawa}  \& {Rodr{\'{\i}}guez del
  Pino}}{{Ruggiero} et~al.}{2019}]{Ruggiero2019}
{Ruggiero} R.,  {Machado} R.~E.~G.,  {Roman-Oliveira} F.~V.,  {Chies-Santos}
  A.~L.,  {Lima Neto} G.~B.,  {Doubrawa} L.,   {Rodr{\'{\i}}guez del Pino} B.,
  2019, \mn@doi [\mnras] {10.1093/mnras/sty3422}, \href
  {http://adsabs.harvard.edu/abs/2019MNRAS.484..906R} {484, 906}

\bibitem[\protect\citeauthoryear{{Shimwell} et~al.,}{{Shimwell}
  et~al.}{2016}]{2016Shimwell}
{Shimwell} T.~W.,  et~al., 2016, \mn@doi [\mnras] {10.1093/mnras/stw661}, \href
  {https://ui.adsabs.harvard.edu/abs/2016MNRAS.459..277S} {459, 277}

\bibitem[\protect\citeauthoryear{{Springel}}{{Springel}}{2005}]{Springel2005}
{Springel} V.,  2005, \mn@doi [\mnras] {10.1111/j.1365-2966.2005.09655.x},
  \href {http://adsabs.harvard.edu/abs/2005MNRAS.364.1105S} {364, 1105}

\bibitem[\protect\citeauthoryear{{Springel} \& {Farrar}}{{Springel} \&
  {Farrar}}{2007}]{Springel2007}
{Springel} V.,  {Farrar} G.~R.,  2007, \mn@doi [\mnras]
  {10.1111/j.1365-2966.2007.12159.x}, \href
  {http://adsabs.harvard.edu/abs/2007MNRAS.380..911S} {380, 911}

\bibitem[\protect\citeauthoryear{{Springel}, {Di Matteo}  \&
  {Hernquist}}{{Springel} et~al.}{2005}]{2005SpringeletalEQ3}
{Springel} V.,  {Di Matteo} T.,   {Hernquist} L.,  2005, \mn@doi [\mnras]
  {10.1111/j.1365-2966.2005.09238.x}, \href
  {https://ui.adsabs.harvard.edu/abs/2005MNRAS.361..776S} {361, 776}

\bibitem[\protect\citeauthoryear{{Turk}, {Smith}, {Oishi}, {Skory}, {Skillman},
  {Abel}  \& {Norman}}{{Turk} et~al.}{2011}]{2011Turk}
{Turk} M.~J.,  {Smith} B.~D.,  {Oishi} J.~S.,  {Skory} S.,  {Skillman} S.~W.,
  {Abel} T.,   {Norman} M.~L.,  2011, \mn@doi [\apjs]
  {10.1088/0067-0049/192/1/9}, \href
  {https://ui.adsabs.harvard.edu/abs/2011ApJS..192....9T} {192, 9}

\bibitem[\protect\citeauthoryear{{White}}{{White}}{2000}]{White2000}
{White} D.~A.,  2000, \mn@doi [\mnras] {10.1046/j.1365-8711.2000.03163.x},
  \href {https://ui.adsabs.harvard.edu/abs/2000MNRAS.312..663W} {312, 663}

\bibitem[\protect\citeauthoryear{{Zhang}, {Yu}  \& {Lu}}{{Zhang}
  et~al.}{2015}]{2015AZhangetal}
{Zhang} C.,  {Yu} Q.,   {Lu} Y.,  2015, \mn@doi [\apj]
  {10.1088/0004-637X/813/2/129}, \href
  {https://ui.adsabs.harvard.edu/abs/2015ApJ...813..129Z} {813, 129}

\bibitem[\protect\citeauthoryear{{ZuHone}}{{ZuHone}}{2011}]{ZuHone2011}
{ZuHone} J.~A.,  2011, \mn@doi [\apj] {10.1088/0004-637X/728/1/54}, \href
  {http://adsabs.harvard.edu/abs/2011ApJ...728...54Z} {728, 54}

\bibitem[\protect\citeauthoryear{{ZuHone}, {Biffi}, {Hallman}, {Rand all},
  {Foster}  \& {Schmid}}{{ZuHone} et~al.}{2014}]{zuhone2014}
{ZuHone} J.~A.,  {Biffi} V.,  {Hallman} E.~J.,  {Rand all} S.~W.,  {Foster}
  A.~R.,   {Schmid} C.,  2014, arXiv e-prints, \href
  {https://ui.adsabs.harvard.edu/abs/2014arXiv1407.1783Z} {p. arXiv:1407.1783}

\bibitem[\protect\citeauthoryear{{van Weeren}, {R{\"o}ttgering}, {Br{\"u}ggen}
  \& {Hoeft}}{{van Weeren} et~al.}{2010}]{2010Vanweeren}
{van Weeren} R.~J.,  {R{\"o}ttgering} H. J.~A.,  {Br{\"u}ggen} M.,   {Hoeft}
  M.,  2010, \mn@doi [Science] {10.1126/science.1194293}, \href
  {https://ui.adsabs.harvard.edu/abs/2010Sci...330..347V} {330, 347}

\makeatother
\end{thebibliography}

\bsp
\label{lastpage}
\end{document}